\documentclass[times,twocolumn]{aastex63}

\usepackage{CJK}
\usepackage{graphicx}
\usepackage{booktabs}

\newcommand{\hi}{\text{HI}}

\newcommand{\ha}{\ifmmode {\rm H}\alpha \else H$\alpha$\fi}
\newcommand{\hb}{\ifmmode {\rm H}\beta \else H$\beta$\fi}
\newcommand{\oiii}{[\textrm{O}~\textsc{III}]}

\newcommand{\nii}{[\textrm{N}~\textsc{II}]}

\newcommand{\vasym}{v_{\text{asym}}}
\newcommand{\vasymbar}{\overline{v}_{\text{asym}}}
\newcommand{\logv}{\log \overline{v}_{\text{asym}}}
\newcommand{\dlogv}{\Delta \log \overline{v}_{\text{asym}}}

\newcommand{\vab}{v_{a_1/b_1}}
\newcommand{\vabbar}{\overline{v}_{a_1/b_1}}
\newcommand{\logab}{\log \overline{v}_{a_1/b_1}}
\newcommand{\dlogab}{\Delta \log \overline{v}_{a_1/b_1}}

\newcommand{\logm}{\log M_\star}

\newcommand{\dsfr}{\Delta \log \mathrm{SFR}}

\newcommand{\doh}{\Delta \log(\mathrm{O/H})_\mathrm{Re}}
\newcommand{\fhi}{f_{\mathrm{\hi}}}
\newcommand{\dfhi}{\Delta f_{\mathrm{\hi}}}

\newcommand{\kms}{\ \text{km}\ \text{s}^{-1}}
\newcommand{\pdis}{d_{\text{p}}}

\received{Sep 1, 2021}
\revised{Oct 10, 2021}
\accepted{\today}
\submitjournal{ApJ}

\shorttitle{Velocity Map Morphology, Star Formation and Metallicity}
\shortauthors{Feng et al.}

\begin{document}
\begin{CJK*}{UTF8}{gbsn}

\title{The Velocity Map Asymmetry of Ionized Gas in MaNGA \\ II. Correlation between Velocity Map Morphology, Star Formation, and Metallicity in Regular Disk Galaxies}

\correspondingauthor{Shuai Feng}
\email{sfeng@hebtu.edu.cn}

\author[0000-0002-9767-9237]{Shuai Feng (冯帅)}
\affiliation{College of Physics, Hebei Normal University, 20 South Erhuan Road, Shijiazhuang 050024, China}
\affiliation{Shijiazhuang Key Laboratory of Astronomy and Space Science, Shijiazhuang 050024, China}
\affiliation{Hebei Key Laboratory of Photophysics Research and Application, Shijiazhuang 050024, China}

\author[0000-0002-3073-5871]{Shiyin Shen (沈世银)}
\affiliation{Shanghai Astronomical Observatory, Chinese Academy of Sciences, 80 Nandan Road, Shanghai 200030, China}
\affiliation{Shanghai Key Lab for Astrophysics, Shanghai 200234, China}

\author[0000-0003-3226-031X]{Yanmei Chen (陈燕梅)}
\affiliation{School of Astronomy and Space Science, Nanjing University, Nanjing 210093, Peopleʼs Republic of China}
\affiliation{Key Laboratory of Modern Astronomy and Astrophysics (Nanjing University), Ministry of Education, Nanjing 210093, Peopleʼs Republic of China}
\affiliation{Collaborative Innovation Center of Modern Astronomy and Space Exploration, Nanjing 210093, Peopleʼs Republic of China}

\author[0000-0002-7928-416X]{Y. Sophia Dai (戴昱)}
\affiliation{Chinese Academy of Sciences South America Center for Astronomy (CASSACA) / National Astronomical Observatories of China (NAOC), \\ 20A Datun Road, Beijing 100012, Peopleʼs Republic of China}

\author[0000-0002-4499-1956]{Jun Yin (尹君)}
\affiliation{Shanghai Astronomical Observatory, Chinese Academy of Sciences, 80 Nandan Road, Shanghai 200030, China}

\author[0000-0003-1359-9908]{Wenyuan Cui (崔文元)}
\affiliation{College of Physics, Hebei Normal University, 20 South Erhuan Road, Shijiazhuang 050024, China}
\affiliation{Shijiazhuang Key Laboratory of Astronomy and Space Science, Shijiazhuang 050024, China}

\author[0000-0002-5815-2387]{Mengting Ju (居梦婷)}
\affiliation{School of Astronomy and Space Science, University of Chinese Academy of Sciences (UCAS), Beijing 100049, China}

\author[0000-0003-1454-2268]{Linlin Li (李林林)}
\affiliation{College of Physics, Hebei Normal University, 20 South Erhuan Road, Shijiazhuang 050024, China}
\affiliation{Shijiazhuang Key Laboratory of Astronomy and Space Science, Shijiazhuang 050024, China}

\begin{abstract}
The morphology of ionized gas velocity maps provides a direct probe of the internal gas kinematics of galaxies. Using integral field spectroscopy from SDSS-IV MaNGA, we analyze a sample of $528$ low-inclination, regular disk galaxies to investigate the correlations between velocity map morphology, star formation rate, and gas-phase metallicity. We quantify velocity map morphology using harmonic expansion and adopt two complementary diagnostics: the global kinematic asymmetry, which traces non-axisymmetric perturbations, and the first-order term ratio, which captures axisymmetric radial motions. We find that galaxies with higher kinematic asymmetry are more likely to deviate from the scaling relations, typically lying either above or below the star formation main sequence and systematically below the mass-metallicity relation. In contrast, the first-order term ratio shows only a correlation with gas-phase metallicity in the low-mass range and no significant dependence on star formation rate. Moreover, galaxies below the mass-metallicity relation generally exhibit higher HI gas fractions. These results suggest that external gas accretion is the primary driver of the observed phenomena: inflowing metal-poor gas increases velocity map asymmetry in disk galaxies, dilutes the metallicity, and triggers enhanced star formation. Feedback-driven outflows, bar- and spiral-driven inflows, and galaxy mergers may also contribute, but likely play a secondary role.
\end{abstract}


\section{Introduction} \label{sec:intro}

Velocity maps describe the line-of-sight velocities across galaxies, providing powerful diagnostics of galactic structure and dynamical evolution. By tracing the motion of baryonic matter, they retain imprints of a galaxy's dynamical history and ongoing processes. Since such motions govern the redistribution of mass and angular momentum, velocity maps also offer insight into mechanisms that will shape future evolution. In particular, ionized gas velocity maps, derived from optical emission lines, encode both large-scale gas motions and small-scale dynamical features, such as regular rotation, inflows, outflows, and turbulence \citep[e.g.,][]{Sofue2001, Veilleux2005, Glazebrook2013, ForsterSchreiber2020}. With the advent of integral field spectroscopy (IFS) surveys, such as CALIFA \citep{Sanchez2012}, SAMI \citep{Bryant2015}, and MaNGA \citep{Bundy2015}, high-quality ionized gas velocity maps can now be obtained for statistically significant samples, enabling robust studies of galaxy kinematics.

The morphology of a galaxy's velocity map provides a direct and intuitive view of its internal motions. In an idealized rotating disk, the velocity map exhibits a symmetric spider-like pattern, where the rotation axis aligns with the direction of the zero-velocity line \citep{vanderKruit1978, DiTeodoro2015}. The kinematic position angle, on the other hand, is defined along the line of maximum velocities, which is $90^{\circ}$ away from the rotation axis. In the real universe, however, disk galaxies rarely conform perfectly to this simple model. Many systems show at least some degree of departure from axisymmetric rotation, reflecting the presence of non-circular motions driven by external perturbations or internal dynamical instabilities \citep{ForsterSchreiber2009, GarciaLorenzo2015}. Investigating these departures is therefore essential for developing a kinematic understanding of the physical processes that influence galaxy evolution, including gas accretion, stellar and AGN feedback, and gravitational interactions.

The first step in investigating deviations from regular rotation is to identify them in observed velocity maps. A straightforward method is visual inspection, which efficiently identifies obvious departures from the ideal rotating-disk pattern \citep[e.g.][]{Flores2006,Law2009}. However, this approach lacks sensitivity to subtle features and does not provide a quantitative description. A more systematic strategy involves fitting a simple rotating-disk model to the velocity map and interpreting the residuals as signatures of non-circular motions \citep[e.g.][]{Fathi2006, Genzel2011}. Yet such residuals are often too diverse in morphology to be easily parameterized, which limits their usefulness for large statistical samples. When the type of non-circular motion is known a priori, tailored kinematic models, such as those designed for bars or radial flows, can be employed \citep[e.g.][]{Spekkens2007, DiTeodoro2021}, but these models are intrinsically restricted to specific motion types. In contrast, harmonic decomposition offers a model-independent framework for quantifying the morphology of velocity maps. By expanding the velocity map into a Fourier series, this method naturally separates the circular component from non-circular features \citep[e.g.][]{Schoenmakers1997,Krajnovic2006}. Because of its flexibility and general applicability, harmonic decomposition has become a powerful tool for the statistical analysis of large galaxy samples.

Building on the strengths of harmonic decomposition, numerous observational studies have employed this technique to quantify deviations from regular disk rotation and to characterize asymmetric features in galaxy velocity maps \citep[e.g.,][]{Shapiro2008, Bloom2017, Feng2022}. These studies indicate that certain types of galaxies are more likely to exhibit peculiar velocity map morphologies, suggesting that distinct physical processes operating in these systems have a strong impact on gas kinematics. For instance, strong tidal forces in interacting or merging systems often induce strong non-circular components, producing highly asymmetric velocity maps \citep[e.g.][]{Hung2016, Bloom2017}. Consequently, harmonic analysis of the velocity map has become a widely used diagnostic tool for identifying galaxy interactions and mergers \citep{Shapiro2008, Feng2020}. Likewise, barred spiral galaxies commonly exhibit enhanced higher-order harmonic terms in regions aligned with the stellar bar, consistent with bar-driven perturbations that substantially modify local gas kinematics \citep[e.g.][]{Schoenmakers1997, Wong2004, Spekkens2007}. In addition, AGN host galaxies frequently display complex, non-axisymmetric velocity structures, likely reflecting the dynamical influence of AGN-driven outflows on the surrounding interstellar medium \citep[e.g.][]{MullerSanchez2011, Venturi2018}.

Recent results from large IFS surveys have shown that non-circular motions in ionized gas velocity maps are not confined to extreme systems such as mergers, barred galaxies, or AGN hosts. Instead, irregular and asymmetric kinematic features are commonly observed across the full diversity of galaxy populations \citep[e.g.][]{Ho2016, Bloom2018, Feng2022}, suggesting that departures from idealized rotating disk kinematics are a widespread feature of galaxies. Within the cosmic ecosystem framework, baryon cycling between galactic disks, the circumgalactic medium (CGM), and the intergalactic medium (IGM) is ubiquitous and accompanies the evolution of all galaxies \citep{Naab2017,Somerville2015, Tumlinson2017}. Such exchanges can manifest kinematically as departures from idealized circular motion, including inflows, outflows, and radial redistribution within the disk, thereby shaping both the short-term regulation of star formation and the long-term pathways of galactic evolution.

A natural question that follows is whether such velocity map irregularities are systematically connected to key global properties that also reflect baryon cycling. The star formation rate (SFR) and gas-phase metallicity are widely used as primary diagnostics of a galaxy's evolutionary state, tracing both its current baryonic processes and its integrated growth history \citep[e.g.][]{Brinchmann2004, Tremonti2004, Finlator2008, Mannucci2010, Dave2011, Lilly2013}. If the morphology of velocity maps is shaped by mechanisms such as gas accretion, stellar or AGN feedback, or internal dynamical instabilities (e.g. bars or spiral perturbations), then measurable imprints on SFR and metallicity should be expected. Motivated by this framework, we conduct a statistical analysis of the relationships between velocity map asymmetry, SFR, and metallicity using a large sample of galaxies with spatially resolved spectroscopic and kinematic measurements.

This paper is organized as follows. Section \ref{sec:data} introduces the galaxy sample and IFS data, and describes the parameterization of velocity map morphology. Section \ref{sec:result} analyzes the correlations between velocity map morphology, SFR, and metallicity. In Section \ref{sec:dis}, we discuss the kinematic picture implied by velocity map morphology and explore the physical origins of their connections with galaxy properties. Finally, Section~\ref{sec:sum} provides a brief summary. Throughout this paper, we adopt a cosmological model with $h=0.7$, $\Omega_{\mathrm{m}}=0.3$, and $\Omega_{\Lambda}=0.7$.

\section{Data}\label{sec:data}

\subsection{MaNGA Survey}

The Mapping Nearby Galaxies at Apache Point Observatory (MaNGA) survey \citep{Bundy2015} is one of the three core programs of SDSS-IV \citep{Blanton2017}. Its goal is to obtain spatially resolved spectroscopic data using an integral field unit (IFU) mounted on the 2.5-meter Sloan Foundation Telescope at Apache Point Observatory \citep{Smee2013, Drory2015, Gunn2006}. The survey targets galaxies spanning a broad range of stellar masses and colors \citep{Yan2016b, Wake2017} within the redshift range $0.01 < z < 0.15$, providing a representative sample for studying the general properties of galaxies in the local universe.

All MaNGA data products were made publicly available in the final data release of SDSS-IV \citep[SDSS DR17;][]{SDSSDR17}, which includes observations of approximately 10,000 unique galaxies. The raw data are processed using the MaNGA Data Reduction Pipeline (DRP; \citealt{Law2015, Yan2016a}) and further analyzed with the MaNGA Data Analysis Pipeline (DAP; \citealt{Westfall2019, Belfiore2019}). The DAP products provide two-dimensional maps of key physical parameters, including line-of-sight velocity and the integrated fluxes of various emission lines, which are directly suitable for scientific analysis. In addition, an independent set of data products is available from PIPE3D \citep{Sanchez2022}, offering complementary measurements for comparison and cross-validation.

\subsection{Parametric Characterization of Velocity Maps}\label{sec:fit_method}

\begin{figure*}
    \centering
    \includegraphics[width=\textwidth]{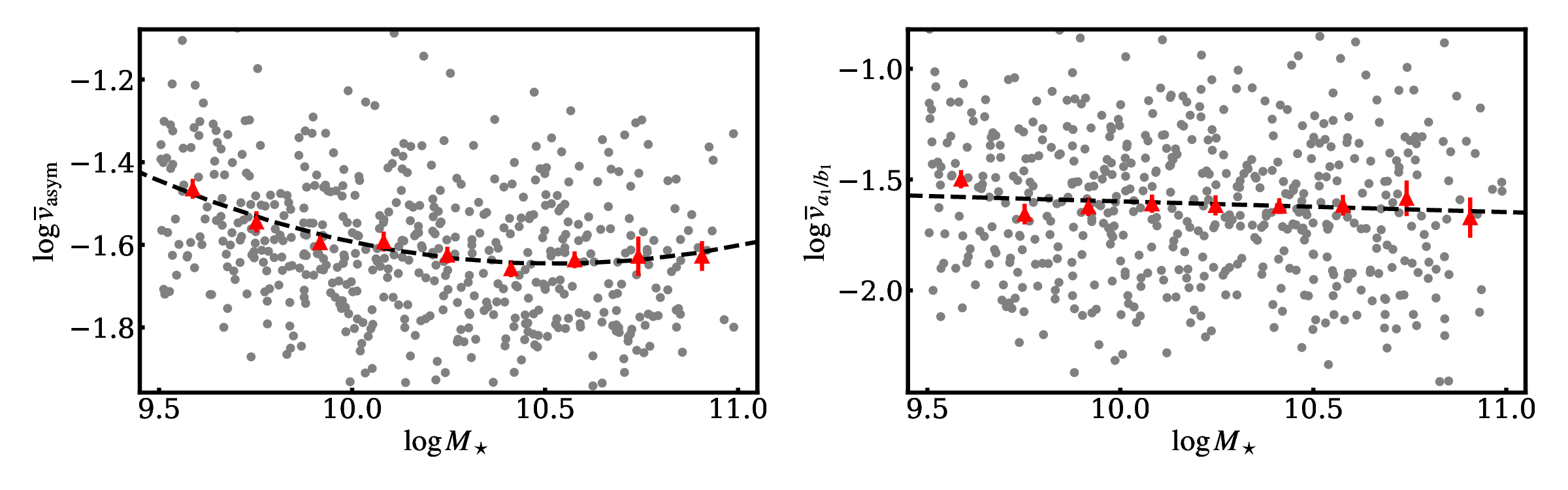}
    \caption{The relationship between stellar mass and velocity map morphology (\textit{Left}: kinematic asymmetry $\logv$. \textit{Right}: first-order term ratio $\logab$). The red triangles represent the median value of velocity map morphology parameters within the given stellar mass bin. The dashed line is a linear fit to these median values as a function of stellar mass. }
    \label{fig:logm_logv}
\end{figure*}

We analyze the ionized gas velocity maps based on the $\mathrm{H}\alpha$ velocity maps from the MaNGA \textsc{DAP}. We adopt the \textsc{HYB10} analysis scheme, in which stellar kinematics are derived from binned spectra via Voronoi tessellation, while emission-line measurements are performed on individual spaxels. In the \textsc{DAP} fitting process, the velocities of all emission lines are tied together, so the $\mathrm{H}\alpha$ velocity map reflects the common kinematic solution constrained by all detected lines. This approach preserves the native spatial resolution, making the maps well-suited for detailed kinematic analysis.

For each galaxy, the line-of-sight velocity along an elliptical ring with semimajor axis $R$ and azimuthal angle $\psi$ in the galaxy plane can be expressed by the harmonic expansion \citep{Franx1994, Schoenmakers1997, Krajnovic2006}, 
\begin{equation}
    V(R,\psi) = a_0 + \sum_{n=1}^{N} \big[\,a_n\,\sin(n\psi) + b_n\,\cos(n\psi)\,\big],
\end{equation}
where $a_0$ is the systemic velocity, $b_1$ describes the dominant circular rotation, and $a_1$ measures departures from pure rotation along the kinematic minor axis, typically associated with axisymmetric radial flows \citep{DiTeodoro2021, Genzel2023}. Terms of order $n \geq 2$ capture progressively more complex perturbations to the velocity map, such as bar-driven elliptical streaming, spiral-arm streaming motions, and warps \citep{Schoenmakers1997, Wong2004, Spekkens2007}. \footnote{In this work, `circular motions' refers specifically to the $b_1$ term, while `non-circular motions' encompass $a_1$ as well as all higher-order terms ($n \geq 2$).} A compact but equivalent form uses the amplitude and phase parameters:
\begin{equation}
    V(R, \psi) = a_0 + \sum_{n=1}^{N} k_n \cos\!\big[n(\psi - \phi_n)\big],
\end{equation}
where
\begin{equation}
    k_n = \sqrt{a_n^2 + b_n^2}
\end{equation}
\begin{equation}
    \phi_n = \mathrm{arctan}\frac{a_n}{b_n}.
\end{equation}

We adopt two metrics to quantify deviations of observed velocity maps from an ideal rotating-disk morphology. The first is the kinematic asymmetry, a widely used metric that measures the relative strength of non-circular motions compared to ordered rotation \citep[e.g.,][]{Krajnovic2006, Shapiro2008, Feng2022}. Two normalization schemes have been proposed in the literature: one based on the total first-order amplitude $k_1$ \citep[e.g.,][]{Bloom2017, Feng2022}, and the other on $|b_1|$ \citep[e.g.,][]{Shapiro2008}, which more directly traces the circular rotation. We adopt the $|b_1|$-normalized form:
\begin{equation}\label{eq:vasym}
    \vasym = \frac{k_2 + k_3 + k_4 + k_5}{4|b_1|}.
\end{equation}
In this work, we truncate the harmonic expansion at $n=5$, as higher-order terms ($n>5$) contribute negligibly and do not affect our results.

The second metric, first-order term ratio $\vab$, measures the relative amplitude of the first-order sine term compared to the rotational (cosine) term in the harmonic expansion \citep[e.g.,][]{Wong2004, Fathi2006, Genzel2023}, 
\begin{equation}
    \vab = \frac{|a_1|}{|b_1|}.
\end{equation}
This ratio is particularly sensitive to velocity components along the minor axis, such as those induced by radial inflows or outflows. Whereas $\vasym$ traces higher-order non-circular motions ($n \geq 2$), $\vab$ provides a complementary diagnostic that captures first-order deviations from pure rotation. 

We employ the \texttt{kinemetry} package \citep{Krajnovic2006} to fit the ionized gas velocity maps and derive the values of harmonic terms. Each velocity map is decomposed into concentric elliptical rings with a fixed width of $1\arcsec$. The fitting is performed in two steps: in the first step, the position angle and ellipticity are allowed to vary; in the second step, they are fixed to the average values measured between $0.5R_e$ and $1.5R_e$.

Based on the second-step results, we compute the kinematic asymmetry parameter $\vasym$ and the first-order term ratio $\vab$ for each elliptical ring. The global values for each velocity map, denoted as $\vasymbar$ and $\vabbar$, are defined as the median of these quantities within one effective radius ($R_e$). Since both parameters are generally smaller than unity, we adopt $\logv$ and $\logab$ in the subsequent statistical analysis to improve the dynamic range and facilitate comparison across the sample.

To estimate uncertainties, we generate $200$ Monte Carlo realizations of each galaxy's velocity map. Each Monte Carlo realization is generated by perturbing the velocity map spaxel-by-spaxel according to Gaussian noise with the DAP velocity and uncertainty as the mean and standard deviation. The same two-step kinemetric fitting is applied to each mock velocity map. The final estimates of $\logv$ and $\logab$ are taken as the median over the $200$ realizations, with their standard deviations providing the associated uncertainties. 

\subsection{Sample Selection}

We select galaxies from the MaNGA survey with specific star formation rates (sSFR) higher than $\log(\mathrm{SFR}/M_\star) = -11.5$, ensuring sufficiently strong $\ha$ emission for reliable measurements of velocity map morphology. Stellar masses and star formation rates are adopted from the MPA-JHU catalog \citep{Brinchmann2004, Kauffmann2003b}.

To focus on disk galaxies, we select those with Sersic indices $n_s < 2.5$ and stellar masses $9.5 < \logm < 11.0$, where the Sersic indices are taken from \citet{Simard2011}. Since the morphology of the velocity map is known to correlate with the inclination of the galaxy \citep[e.g.,][]{Feng2022}, we further restrict the sample to galaxies with axis ratios $b/a > 0.5$ to reduce projection effects.

To avoid contamination from galaxies whose kinematic asymmetries are likely dominated by well-understood processes, we exclude three categories: mergers, barred galaxies, and AGN hosts. For mergers, we first remove objects flagged by the image-based machine-learning classifications of \citet{Walmsley2023}, defined as $\texttt{merging\_none\_fraction} \leq 0.5$, and then exclude galaxies with bright ($r < 17.77$) companions within a projected distance of $\pdis < 150\ \mathrm{kpc}$ and a line-of-sight velocity difference $|\Delta v| < 500 \kms$ \citep{Feng2019, Feng2020}. Barred galaxies are removed based on the $\texttt{bar\_no\_fraction} \leq 0.5$ according to \citet{Walmsley2023}. To exclude AGN, we retain only BPT-classified star-forming galaxies \citep{Kauffmann2003a, Kewley2006}, based on the $\nii/\ha$ versus $\oiii/\hb$ diagnostic diagram using nuclear fluxes measured within a $2.5\arcsec$ aperture.

After applying all selection criteria, the final sample consists of $528$ galaxies.

\section{Result} \label{sec:result}

\subsection{Correlation between Velocity Map Morphology and Stellar Mass}

Figure~\ref{fig:logm_logv} displays the distribution of the $528$ selected low-inclination regular disk galaxies in the $\logm$--$\logv$ (left panel) and $\logm$--$\logab$ (right panel) planes. In both panels, red triangles represent the median values in stellar mass bins. The black dashed lines in each panel represent the best-fit linear relations to the data in the respective planes.

The results shown in Figure~\ref{fig:logm_logv} reveal that the two parameters used to characterize the velocity map morphology exhibit different dependencies on stellar mass in regular disk galaxies. Specifically, the value of $\logv$ increases significantly toward lower stellar masses, consistent with previous studies \citep[e.g.,][]{Bloom2017, Feng2022}. In contrast, $\logab$ exhibits little correlation with stellar mass. This may suggest that non-circular motions associated with $\logab$ may be largely independent of stellar mass. On the other hand, the mass dependence of $\logv$ may indicate that non-circular motions captured by this parameter are more prominent in lower-mass galaxies.

Given the observed dependence of velocity map morphology on stellar mass, we define residual parameters for both $\logv$ and $\logab$ after removing their stellar-mass trends. These residuals quantify the deviation of each parameter from its expected value at fixed stellar mass, allowing a fair comparison of velocity map morphology across galaxies of different masses while preserving the diagnostic power of the original quantities.

For $\logv$, the residual is
\begin{equation}
    \dlogv = \logv - \log \overline{v}_{\mathrm{asym},0},
\end{equation}
where the baseline is given by the best-fit quadratic relation
\begin{equation}
    \log \overline{v}_{\mathrm{asym},0} = 0.192\,x^2 - 0.201\,x - 1.593,
\end{equation}
with $x = \logm - 10$.

Although $\logab$ shows little correlation with stellar mass, we apply the same procedure:
\begin{equation}
    \dlogab = \logab - \log \overline{v}_{a_1/b_1,0},
\end{equation}
where
\begin{equation}
    \log \overline{v}_{a_1/b_1,0} = -0.048\,x - 1.599.
\end{equation}

In the subsequent analysis, these residuals serve as our primary indicators for characterizing the morphology of velocity maps across galaxies.

\subsection{Correlation between Velocity Map Morphology and Star Formation Rate}\label{sec:sfms}

\begin{figure*}
    \centering
    \includegraphics[width=\textwidth]{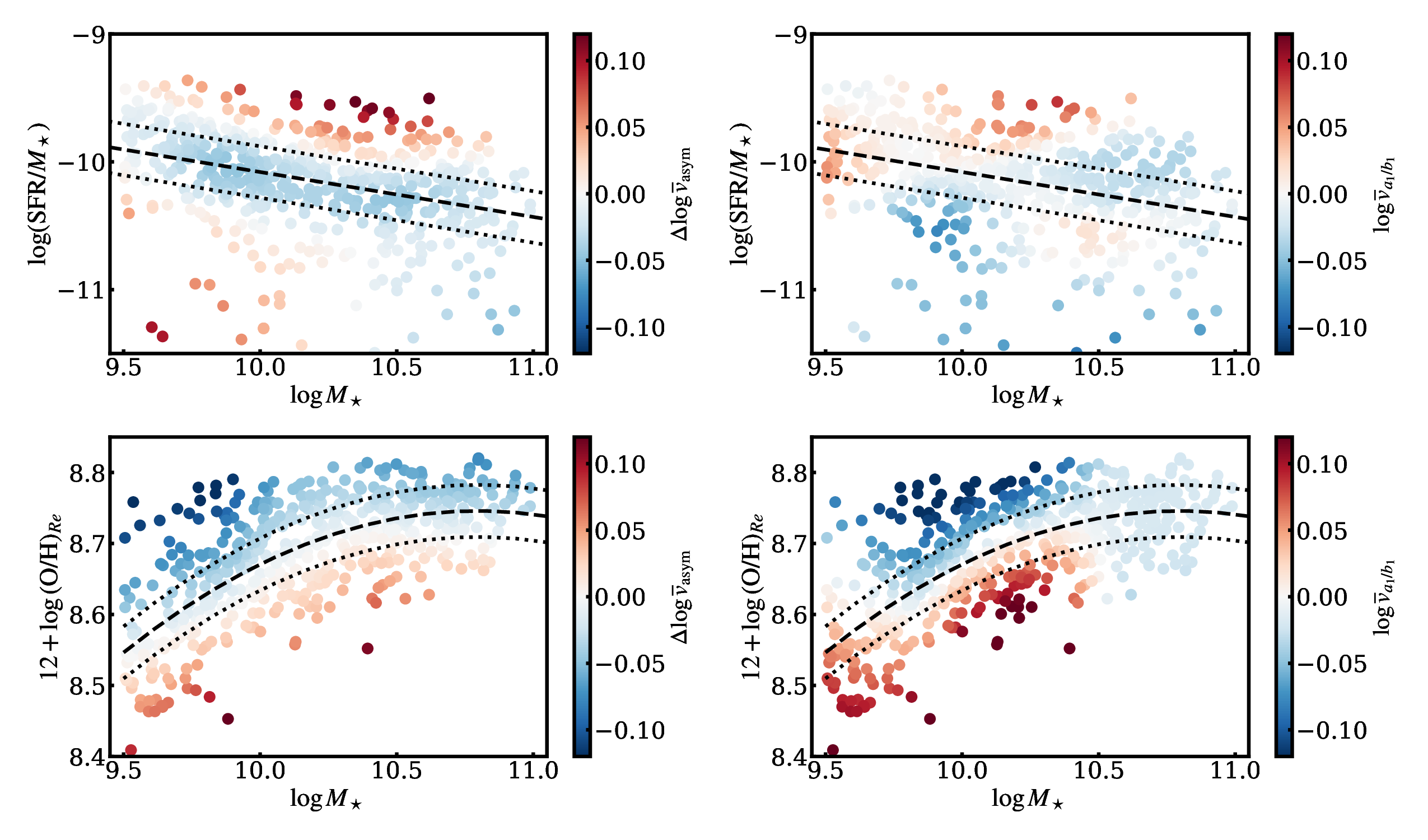}
    \caption{Dependence of the star formation main sequence (SFMS, \textit{top}) and the mass-metallicity relation (MZR, \textit{bottom}) on velocity map morphology. In the \textit{left} panels, colors represent $\dlogv$, while in the \textit{right} panels, colors represent $\dlogab$. The dashed lines mark the best-fitting trend of SFMS or MZR, and the dotted lines indicate the 25th and 75th percentiles. }
    \label{fig:SFMS_MZR}
\end{figure*}

We begin by examining how velocity map morphology varies across the star formation main sequence (SFMS). The top panels of Figure~\ref{fig:SFMS_MZR} show the distribution of galaxies in the sSFR--stellar mass plane. Galaxies are color-coded by $\dlogv$ (top-left) and $\dlogab$ (top-right), with red indicating greater deviation from a regular rotating disk. To characterize the SFMS, we follow the approach of \citet{Salim2007} and fit a linear relation to the median sSFR in stellar mass bins. The resulting best-fit relation is
\begin{equation}\label{eq:sfms}
    \log(\mathrm{SFR}/M_\star) = -0.352x - 10.08,
\end{equation}
where $x = \logm - 10$. This fit is plotted as a black dashed line in Figure~\ref{fig:SFMS_MZR}, with the 25th and 75th percentile boundaries indicated by dotted lines. 

To investigate how velocity map morphology connects to the SFMS, we apply Locally Estimated Scatterplot Smoothing (LOESS), a non-parametric regression method that fits low-order polynomials to localized subsets of the data \citep{Cleveland1979}. This approach enables us to trace the overall variation of both $\dlogv$ and $\dlogab$ across the sSFR--stellar mass plane, thereby providing a clearer view of their correlations with the SFMS. The smoothing is performed using the \texttt{LOESS} package \citep{Cappellari2013}.

In the top-left panel, $\dlogv$ exhibits a clear dependence on SFR at fixed stellar mass. Galaxies located along the SFMS generally have negative values of $\dlogv$, indicating more symmetric velocity maps. By contrast, galaxies above the SFMS tend to show positive values, corresponding to more asymmetric velocity morphologies. Below the SFMS, the behavior is more complex: at lower stellar masses ($\logm<10.2$), galaxies still display relatively negative values, whereas at higher masses the values become comparable to those on the SFMS or even turn positive. These patterns suggest that galaxies deviating from the SFMS often exhibit stronger velocity map asymmetries, while galaxies on the SFMS tend to maintain regular, disk-like rotation.

In contrast, the top-right panel shows that $\dlogab$ has no correlation with position on the SFMS. While localized fluctuations exist, there is no systematic trend either above or below the SFMS. This suggests that the strength of the $|a_1|/|b_1|$ component in the velocity map is largely independent of star formation activity.

To complement the LOESS-based analysis that traces global trends in the median value of velocity map morphology across the SFMS, we also examine the full statistical distributions of these parameters, explicitly accounting for measurement uncertainties. For each galaxy, $200$ Monte Carlo realizations are performed, resulting in $200$ measurements of the velocity morphology parameters that explicitly account for observational noise and fitting uncertainties. All galaxies are grouped according to whether their sSFR lies above the $75$th percentile, within the $25$th-$75$th percentile range, or below the $25$th percentile of the residuals relative to the best-fit SFMS. Using all $200$ realizations per galaxy, we then construct the full distributions of $\dlogv$ and $\dlogab$ for each subset. This approach provides a more robust statistical basis for assessing the relationship between velocity map morphology and star formation activity.

The top-left panel of Figure \ref{fig:dv_SFMS_MZR} shows the cumulative distributions of $\dlogv$ for galaxies above (blue), on (green), and below (red) the SFMS. Galaxies above and below the sequence display markedly different distributions compared to those on the SFMS. Specifically, the two off-sequence subsets contain a larger fraction of galaxies with positive $\dlogv$, whereas galaxies on the SFMS show a higher fraction toward negative values. These results reinforce the trend observed in Figure \ref{fig:SFMS_MZR}, confirming that kinematic asymmetry is statistically correlated with star formation activity. 

In contrast to the $\dlogv$ distributions, the $\dlogab$ values show no significant differences among the three subsets, with overall distribution shapes remaining similar. This finding is consistent with Figure~\ref{fig:SFMS_MZR}, indicating that the strength of the $|a_1|/|b_1|$ velocity component is largely insensitive to star formation activity.

\subsection{Correlation between Velocity Map Morphology and Gas-phase Metallicity}\label{sec:mzr}

\begin{figure*}
    \centering
    \includegraphics[width=\textwidth]{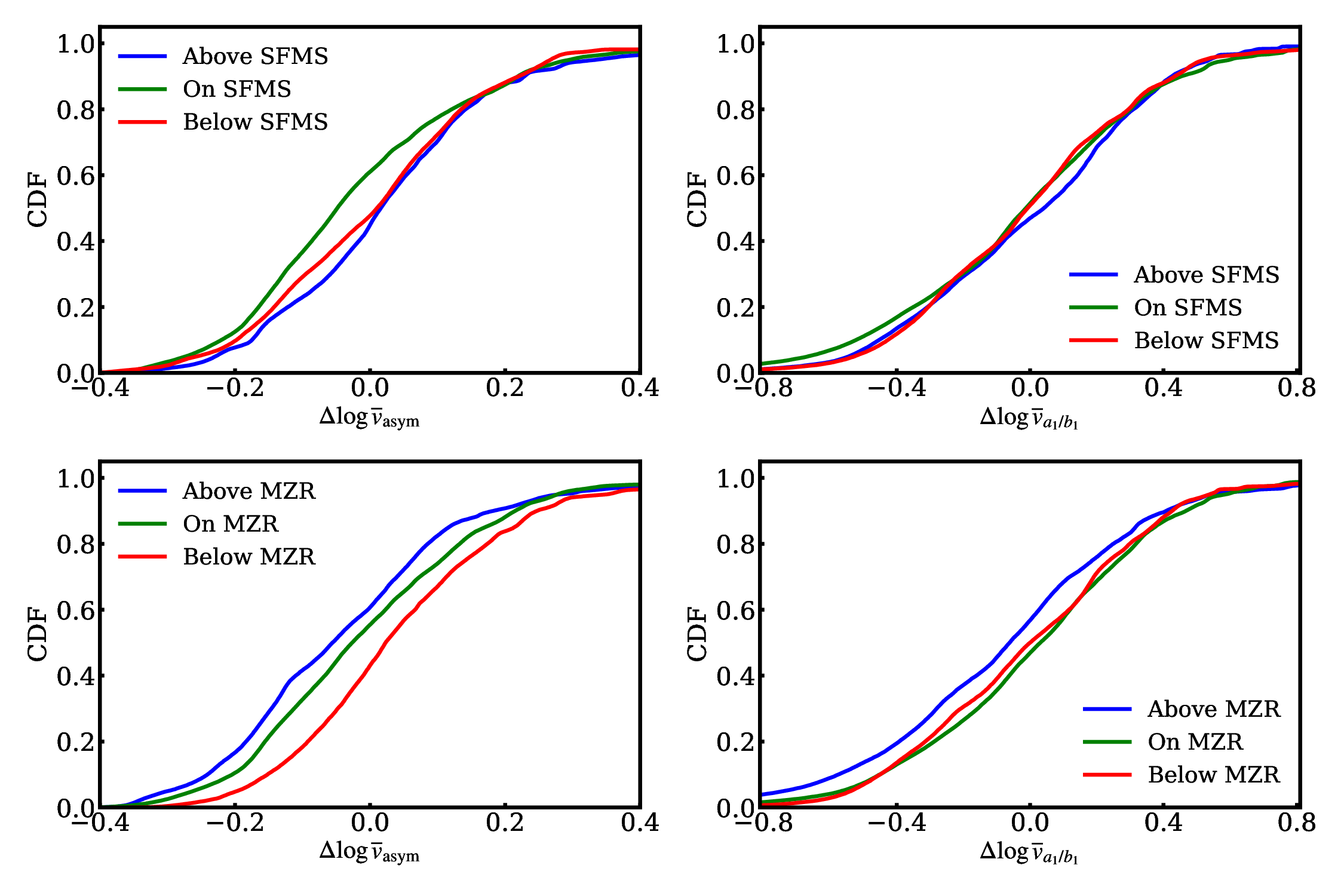}
    \caption{Cumulative distributions of $\dlogv$ (left) and $\dlogab$ (right) for galaxies located at different positions relative to the SFMS (top panels) and the MZR (bottom panels). The cumulative distributions are constructed from the individual distribution functions of $\dlogv$ and $\dlogab$ derived for each galaxy.}
    \label{fig:dv_SFMS_MZR}
\end{figure*}

We next examine how velocity map morphology varies across the mass-metallicity relation (MZR). The bottom-left and bottom-right panels of Figure \ref{fig:SFMS_MZR} show galaxies in the gas-phase metallicity-stellar mass plane. Following \citet{Sanchez2013}, we adopt the gas-phase metallicity at the effective radius as the representative value for each galaxy. These measurements are taken from the PIPE3D catalog \citep{Sanchez2022} and are based on star-forming spaxels using the O3N2 calibration of \citet{PP04}. To parameterize the MZR, we follow the approach of \citet{Tremonti2004} and fit a second-order polynomial to the median metallicity values in stellar mass bins. The best-fit relation is
\begin{equation}\label{eq:mzr}
    12 + \log(\mathrm{O/H}) = -0.119 + 2.566x - 5.102x^2,
\end{equation}
where $x = \logm$. This polynomial is shown as a black dashed line in Figure~\ref{fig:SFMS_MZR}, with the $25$th and $75$th percentile boundaries indicated by dotted lines. Galaxies are color-coded by $\dlogv$ (bottom-left) and $\dlogab$ (bottom-right), with red indicating stronger deviations from regular disk rotation. As in the SFMS analysis, we apply LOESS smoothing to highlight global trends.

In the bottom-left panel, $\dlogv$ is correlated with a galaxy's position relative to the MZR at fixed stellar mass: galaxies above the relation typically have negative values of $\dlogv$, indicating more symmetric velocity maps, while galaxies below the relation generally show positive values, reflecting stronger asymmetries. This trend persists across the stellar mass range of $\log(M_\star/M_\odot) \sim 9.5$ to $11.0$, indicating that galaxies lying below the MZR typically show stronger deviations from regular disk kinematics.

In the bottom-right panel, $\dlogab$ shows a broadly similar trend with respect to the MZR. At lower stellar masses ($\logm < 10.5$), galaxies above the relation typically have negative values, whereas those below the relation tend to show positive values. At higher stellar masses ($\logm > 10.5$), however, $\dlogab$ values are approximately zero, suggesting little dependence on metallicity. These results indicate that low-mass galaxies below the MZR possess stronger $|a_1|/|b_1|$ components in their velocity maps. 

To further examine the connection between gas-phase metallicity and velocity map morphology, we also compare the distribution of both $\logv$ and $\logab$ for three subsets divided according to their position relative to the MZR. The results are shown in the bottom panels of Figure \ref{fig:dv_SFMS_MZR}.  

Overall, the results from the distribution analysis are consistent with the LOESS-based trends shown in Figure~\ref{fig:SFMS_MZR}. For the distribution of $\dlogv$ shown in the bottom left panel, the three subsets exhibit clear differences. Galaxies below the MZR (blue) tend to have a higher fraction of low $\dlogv$ values, while those above the MZR (red) show a larger proportion of galaxies with high $\dlogv$. Galaxies near the MZR (green) show a distribution that lies between the above-MZR and below-MZR populations. These distributions suggest that, at fixed stellar mass, galaxies with lower metallicities tend to exhibit more asymmetric velocity maps.

In contrast, the cumulative distributions of $\dlogab$ for the three subset also show differences, but the distinctions are relatively weaker. The subset of galaxies located above the MZR shows a higher fraction of systems with low $\dlogab$, whereas the subsets on and below the MZR do not exhibit significant differences from each other. This result is consistent with Figure \ref{fig:SFMS_MZR}, where $\dlogab$ shows a strong correlation with metallicity only at $\logm < 10.5$. In addition, we specifically compared the distribution of $\dlogab$ for galaxies with $\logm<10.5$ as a function of their location relative to the MZR. The differences are still less pronounced than those seen in $\logv$. This is mainly because the mean uncertainties of $\logab$ ($\sim 0.06$) are larger than those of $\logv$ ($\sim 0.02$). As a result, although the median value of $\logab$ shows a significant correlation with metallicity, the correlation becomes much weaker once the measurement uncertainties are taken into account. Taken together, these results indicate that the dependence of $\dlogab$ on metallicity is far less significant than that of $\dlogv$.

\subsection{Correlation between Velocity Map Morphology, Star Formation Rate, and Gas-Phase Metallicity}\label{sec:fmr}

\begin{figure*}
    \centering
    \includegraphics[width=\linewidth]{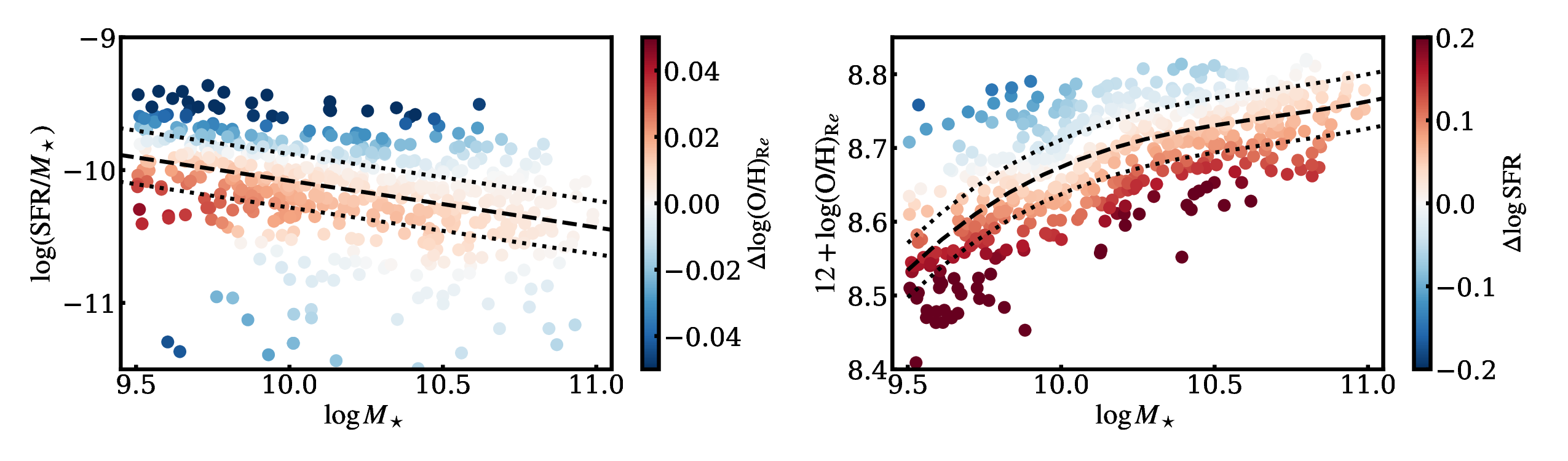}
    \caption{\textit{Left:} galaxies in the stellar mass -- sSFR plane, color-coded by their metallicity offset from the best-fit MZR. \textit{Right:} galaxies in the stellar mass -- metallicity plane, color-coded by their SFR offset from the SFMS.}
    \label{fig:fmr}
\end{figure*}

The previous sections have shown that galaxy velocity map morphology correlates with both star formation activity and gas-phase metallicity. Since SFR and metallicity are themselves strongly linked \citep[e.g.,][]{Mannucci2010}, it is crucial to investigate how the observed kinematic trends depend on these two quantities simultaneously. This section, therefore examines the joint dependence of velocity map morphology on SFR and metallicity to clarify their interconnection. 

We begin with the correlation between SFR and gas-phase metallicity in our sample, as shown in Figure \ref{fig:fmr}. The left panel shows metallicity across the SFMS, with galaxies color-coded by their metallicity offset $\doh$ from the best-fit MZR (Equation \ref{eq:mzr}). The right panel shows the stellar mass-metallicity plane, with galaxies color-coded by their SFR offset $\dsfr$ from the SFMS (Equation \ref{eq:sfms}), highlighting the variation of SFR at fixed stellar mass and metallicity.

The right panel of Figure \ref{fig:fmr} clearly demonstrates an anti-correlation between SFR and gas-phase metallicity at fixed stellar mass, consistent with the well-established fundamental metallicity relation \citep[FMR, ][]{Ellison2008, Mannucci2010, Cresci2019}. In contrast, the trends in the left panel are more complex. For galaxies located on or above the SFMS, $\doh$ shows a monotonic decrease with increasing $\log(\text{SFR}/M_\star)$ at fixed mass, again in agreement with the FMR \citep{Baker2023, Khoram2025}. However, galaxies located $\sim 0.5$ dex below the SFMS exhibit systematically lower metallicities than those on the main sequence, despite having lower SFRs. This discrepancy from previous studies likely arises because samples limited to strong emission-line galaxies generally exclude such low-SFR systems, and hence did not capture this population. 

By comparing Figure \ref{fig:fmr} and Figure \ref{fig:SFMS_MZR}, we find clear correlations among galaxy velocity map morphology, SFR, and gas-phase metallicity. From the left panel of Figure \ref{fig:fmr}, which illustrates the metallicity variation across the SFMS, together with the top panels of Figure \ref{fig:SFMS_MZR} showing how velocity map morphology varies across the SFMS, we infer that galaxies deviating from the SFMS, either above or below, tend to exhibit both more asymmetric velocity maps and lower gas-phase metallicities. Similarly, by comparing the right panel of Figure \ref{fig:fmr}, which displays the variation in SFR with respect to the MZR, with the bottom panels of Figure \ref{fig:SFMS_MZR}, we find that galaxies below the MZR generally show a more asymmetric velocity map and elevated SFRs. Taken together, these results demonstrate that galaxies with disturbed kinematics tend to deviate simultaneously from both the SFMS and the MZR. 

\begin{figure*}
    \centering
    \includegraphics[width=\textwidth]{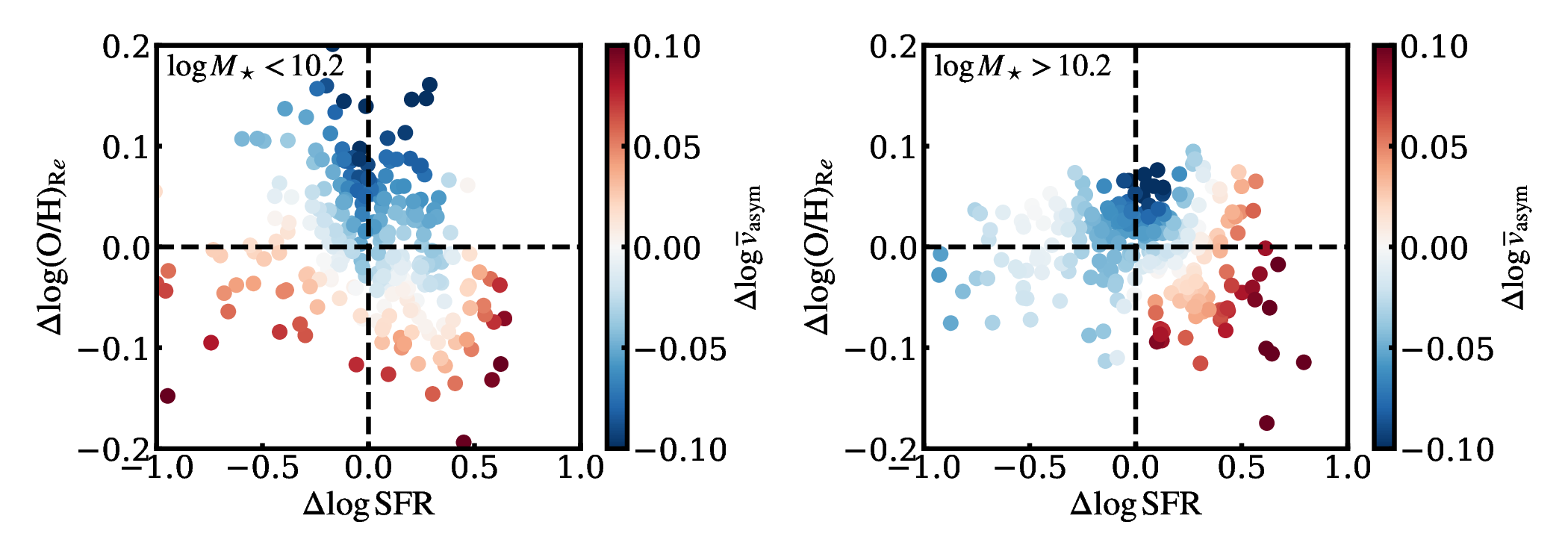}
    \caption{The distribution of the offsets from the SFMS and MZR, where the colors represent $\dlogv$. The \textit{left} and \textit{right} panels display the results of low-mass galaxies ($\logm<10.2$) and high-mass galaxies ($\logm>10.2$), respectively. }
    \label{fig:fmr_dv}
\end{figure*}

To directly probe the interplay among velocity map morphology, SFR, and gas-phase metallicity, we analyze how kinematic asymmetry is distributed across the $\dsfr$--$\doh$ plane. The results are shown in Figure \ref{fig:fmr_dv}, with low-mass galaxies ($\logm < 10.2$) in the left panel and high-mass galaxies ($\logm > 10.2$) in the right.

As expected, galaxies with higher kinematic asymmetry tend to deviate from both the SFMS and the MZR, although the dominant trends vary with stellar mass. For galaxies with $\logm < 10.2$, asymmetric velocity maps are mainly associated with systems below the MZR ($\doh < 0$), regardless of their position relative to the SFMS. This result appears to conflict with Figure \ref{fig:SFMS_MZR}, where galaxies on the SFMS generally display more regular kinematics, suggesting that the kinematic asymmetry of low-mass galaxies might be more strongly linked to metallicity than to SFR. For high-mass galaxies with $\logm > 10.2$, disturbed velocity maps are more often found in systems above the SFMS ($\dsfr > 0$), while their metallicities can lie either on or below the MZR. This trend is broadly consistent with Figure \ref{fig:SFMS_MZR}. Taken together, these results also highlight a mass-dependent behavior: at low stellar mass, velocity map asymmetry is governed primarily by metallicity, whereas the SFR and metallicity play comparable roles at high stellar mass.

\subsection{Correlation between $\hi$ Gas Fraction, Star Formation, and Metallicity}\label{sec:hi}

Cold gas provides the raw material for star formation and plays a central role in regulating both star formation activity and chemical enrichment. Investigating the $\hi$ reservoir is therefore crucial for understanding how galaxy kinematics connect to star formation and metallicity. We examine this connection using $\hi$ mass measurements from the HI-MaNGA catalog \citep{Masters2019, Stark2021}. Among our regular disk sample, $285$ galaxies have reliable $\hi$ detections. 

We use the $\hi$ gas fraction to characterize the atomic gas reservoir of galaxies, defined as
\begin{equation}
    \fhi = \frac{M_{\hi}}{M_{\hi} + M_\star},
\end{equation}
where $M_{\hi}$ and $M_\star$ are the atomic hydrogen mass and stellar masses, respectively. To remove the primary dependence on stellar mass, we further define the residual gas fraction as
\begin{equation}
    \dfhi = \fhi - f_{\text{\hi,0}},
\end{equation}
where the baseline stellar mass -- $\hi$ fraction relation is
\begin{equation}
    f_{\text{\hi,0}} = -0.31 \times (\logm - 10) + 0.385.
\end{equation}
The quantity $\dfhi$ therefore quantifies the relative $\hi$ content of a galaxy at fixed stellar mass.

\begin{figure*}
    \includegraphics[width=\textwidth]{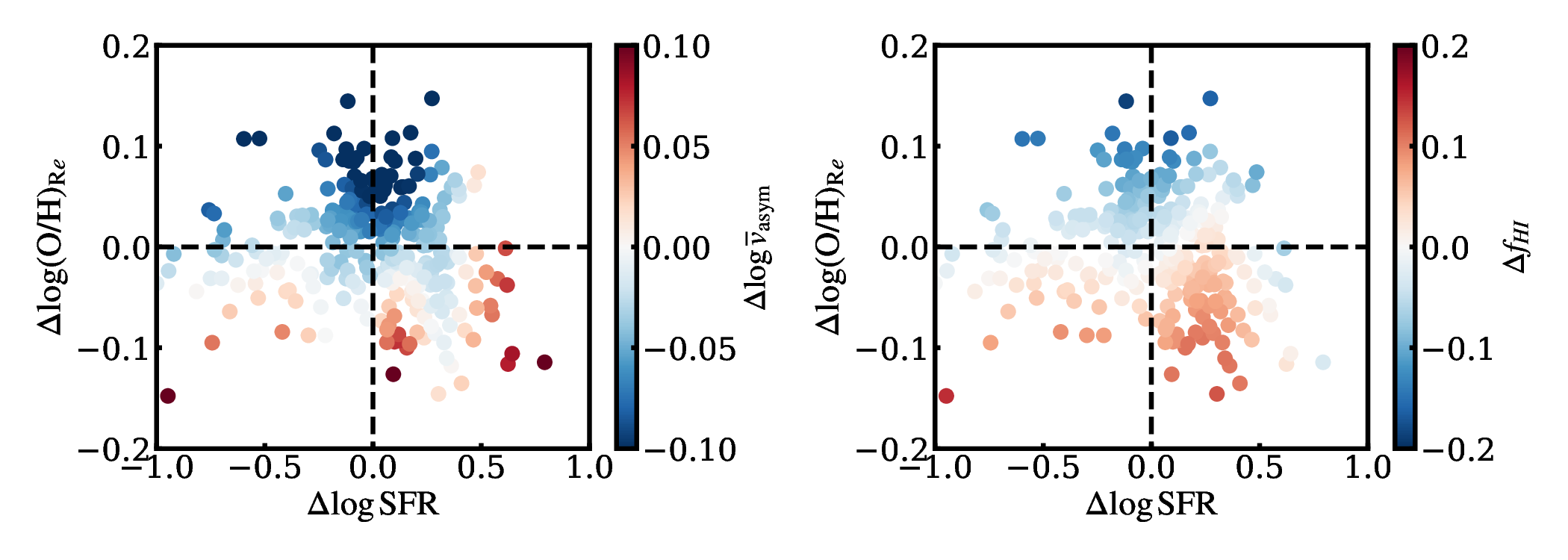}
    \caption{Distribution of velocity map asymmetry and $\hi$ gas content in the $\dsfr$--$\doh$ plane for galaxies with $\hi$ detections. \textit{Left:} colors represent $\dlogv$. \textit{Right:} colors represent the residual $\hi$ gas fraction, $\dfhi$, i.e., the difference in $\hi$ content relative to galaxies of similar stellar mass. }
    \label{fig:fmr_hi}
\end{figure*}

For the subsample of galaxies with $\hi$ detections, we find no direct correlation between $\dfhi$ and $\dlogv$, consistent with the results of \citet{Feng2022}. We therefore focus on how $\dfhi$ relates to SFR and metallicity, and compare these trends with the corresponding dependencies of $\dlogv$. This comparison allows us to explore the potential role of the $\hi$ reservoir. Figure~\ref{fig:fmr_hi} presents these results in the $\dsfr$--$\doh$ plane: the left panel shows the distribution of $\dlogv$ as a function of SFR and metallicity residuals, while the right panel displays the corresponding variation in $\dfhi$.

Although galaxies with $\hi$ detections constitute only about half of the total sample, their kinematic properties exhibit broadly consistent correlations with SFR and metallicity as in the full sample. In particular, galaxies with more asymmetric velocity maps tend to either have enhanced SFRs above the SFMS or reduced gas-phase metallicities below the MZR. The right panel further shows that $\dfhi$ correlates more strongly with metallicity than with SFR: galaxies below the MZR generally have $\dfhi > 0$, while those with SFRs elevated by more than $\sim 0.5$ dex above the SFMS tend to have $\dfhi < 0$ even when lying below the MZR.

These results confirm the expectation that velocity asymmetry does not scale monotonically with cold gas content. Nonetheless, some connections remain. For galaxies with moderate or suppressed SFRs, those with high kinematic asymmetry are more likely also to have higher $\hi$ content and lower metallicities. In contrast, when the SFR is significantly elevated above the SFMS, strong kinematic asymmetry might be associated with reduced $\hi$ content.

\section{Discussion}\label{sec:dis}

\subsection{Impact of Fitting Method} 

As mentioned in previous studies (e.g., \citealt{Schoenmakers1997, Wong2004, Krajnovic2006}), the characterization of velocity map morphology using harmonic expansion depends on several assumptions in the fitting procedure, including the adopted velocity map center, disk inclination, and position angle of the major axis. In this section, we discuss the robustness of our results against variations in these assumptions, focusing on whether adopting alternative assumptions would alter the main conclusions of this work.

In our fiducial analysis, the galaxy center was fixed to the coordinates provided by the MaNGA DAP, which are based on SDSS $r$-band photometry. Since our sample consists mainly of regular disk galaxies, the photometric and kinematic centers are generally well aligned. To evaluate the sensitivity to this choice, we randomly shifted the center by 1-2 pixels ($\sim$ 0.5-1 arcsec) and remeasured the velocity map morphology. After shifting the center, the values of both kinematic asymmetry and the first-order term ratio increase for all galaxies. However, since our analysis focuses on the relative values of these two parameters, the final statistical results remain the same as those obtained with the fiducial assumptions.

For the position angle and inclination, we adopted the values obtained from the kinematic fitting of the velocity map, assuming that they remain constant with radius. Alternatively, some studies use the photometric position angle and inclination, or allow these parameters to vary with radius. We also tested these two alternative assumptions by refitting the velocity maps and repeating our analysis. We found that, for individual galaxies, the values of both kinematic asymmetry and the first-order term ratio differ from the fiducial case. However, the statistical results show no significant differences. Considering that our measurements are limited to the velocity maps within one effective radius, the position angle and inclination of regularly rotating disks are not expected to vary strongly with radius. Moreover, the gas kinematic and photometric orientations of galaxies do not always coincide \citep[e.g.,][]{Jin2016}. Therefore, we consider our fiducial assumption to be more appropriate for the goals of this study. 

\subsection{Kinematic Diagnostics of Non-Circular Gas Motions}

The morphology of gas velocity maps provides a comprehensive description of the internal motions of galaxies, where the degree of deviation from an ideal rotating-disk model reflects the strength of non-circular motions. In this work, we adopt two complementary diagnostics, $\vasym$ and $\vab$, that characterize distinct modes of departure from pure disk rotation.

The ratio $a_1/b_1$ derived from harmonic expansion primarily captures axisymmetric deviations from ordered rotation. A significant departure of $a_1/b_1$ from zero, whether positive or negative, is generally interpreted as evidence for large-scale, nearly axisymmetric radial gas motions, with the sign potentially distinguishing between inflows and outflows when the disk inclination and rotation sense are accurately known (e.g., \citealt{DiTeodoro2021, Genzel2023}). In this work, we adopt $\vab$, defined as the absolute ratio $|a_1|/|b_1|$, which we interpret as the strength of axisymmetric radial flow from the ordered circular rotation. This choice avoids cancellation effects that arise when $a_1/b_1$ changes sign with radius, thereby allowing galaxies with significant but radially varying inflows and outflows to be distinguished from systems with genuinely negligible radial motions.

In contrast, $\vasym$ measures the overall asymmetry of the velocity map and is particularly sensitive to non-axisymmetric perturbations. A wide range of dynamical processes can increase $\vasym$ by disturbing the regular disk rotation. External drivers include gas accretion \citep[e.g.,][]{Stewart2011, Martin2019}, tidal interactions, and mergers \citep[e.g.,][]{Shapiro2008, Feng2020}, while internal mechanisms such as bar- or spiral-induced streaming \citep[e.g.,][]{Wong2004, Spekkens2007} and feedback-driven outflows \citep[e.g.,][]{MullerSanchez2011, Ho2016} may also play a role. In addition, geometric distortions like disk warps can also contribute \citep[e.g.,][]{Wang2023}. Elevated values of $\vasym$ are therefore likely linked to the enhanced influence of these processes.

Although $\vab$ and $\vasym$ emphasize different kinematic modes, they are not entirely independent. Certain dynamical processes, particularly non-axisymmetric but coherent flows such as bar- or spiral-driven streaming, can simultaneously enhance the radial component and excite higher-order harmonics \citep[e.g.,][]{Schoenmakers1997, Wong2004}. As a result, both indicators may increase together, reflecting the fact that some forms of non-circular motion leave multiple imprints on the velocity map. 

\subsection{Physical Origin of the Correlations with Star Formation and Metallicity}

In this section, we interpret the observed correlations between the morphology of the ionized gas velocity map and global galaxy properties. We find that the kinematic asymmetry $\vasym$ shows strong correlations with both star formation rate and gas-phase metallicity across the entire stellar mass range. By contrast, the first-order term ratio $\vab$ only shows correlation with metallicity restricted to galaxies with $\log M_\star < 10.5$, and exhibits no significant dependence on star formation rate. Because non-axisymmetric perturbations can simultaneously elevate the kinematic asymmetry and, to some extent, the first-order term ratio, they are likely the primary drivers of the observed trends, with changes in the radial component reflecting secondary effects of the same processes. 

To place these findings in a physical context, we next consider the mechanisms that can generate non-axisymmetric gas motions and evaluate whether they can account for the observed connections between velocity map morphology, star formation activity, and metallicity. For each mechanism, we discuss the expected kinematic signatures in the velocity map and their potential impact on the chemical and star formation properties of galaxies, and then compare these predictions with our observational results.

\subsubsection{Gas Accretion}

Galaxies can accrete gas from the circumgalactic medium, the intergalactic medium, or along cosmic filaments, thereby supplying fresh fuel for star formation while simultaneously diluting the interstellar metallicity \citep[e.g.,][]{Keres2005, Dekel2006, SanchezAlmeida2014}. Because the angular momentum of the accreted material is often misaligned with that of the host disk and its spatial distribution is typically anisotropic, the inflowing gas tends to settle along non-axisymmetric trajectories \citep[e.g.,][]{Stewart2011, Danovich2015}, enhancing velocity map asymmetries. This mechanism provides a natural explanation for why galaxies with lower metallicities and elevated star formation rates are more likely to exhibit strong kinematic asymmetry. Besides, since such inflows often contain a substantial radial velocity component, they can also increase the first-order term ratio, thereby giving rise to the observed correlation between the first-order term ratio and gas-phase metallicity.

Gas accretion can also explain why some galaxies below the SFMS display both low metallicities and disturbed velocity maps. Cosmological simulations have shown that star-forming galaxies oscillate around the SFMS ridge as a result of repeated cycles of gas compaction, depletion, and replenishment \citep{Tacchella2016, Torrey2018}. Upward departures are driven by fresh accretion, while downturns occur when star formation peaks and outflows suppress further inflow, with subsequent replenishment from the outer disk triggering a new rise. Because the dilution of metallicity and the perturbation of gas kinematics by accretion occur almost instantaneously, whereas a substantial increase in star formation requires the inflowing gas to accumulate and cool, galaxies observed at the onset of an accretion episode may simultaneously show reduced metallicities, asymmetric velocity maps, and relatively low star formation. This interpretation is further supported by our $\hi$ results, which show that galaxies in this phase also tend to exhibit systematically elevated $\hi$ fractions compared to the average population.

The fact that galaxies with low SFR and low metallicity show high kinematic asymmetry only at the low-mass end may be explained by their shallow potential wells or their relatively small $\hi$ reservoirs. When gas accretion first sets in, even a modest inflow can readily disturb the velocity maps of such galaxies, whereas more massive systems are less susceptible to perturbations during the early stages of accretion.

\subsubsection{Gas Outflows}

Stellar feedback and active galactic nuclei activity are known to drive powerful gas outflows that perturb the velocity maps of galaxies \citep[e.g.,][]{Veilleux2005, Hopkins2012}. These winds are often observed as biconical or otherwise asymmetric structures, typically launched along the minor axis, and they introduce velocity components that deviate strongly from ordered disk rotation \citep{MullerSanchez2011, Bouche2012, Ho2014}. When such flows are superimposed on the underlying rotational map, they introduce additional non-circular components that distort the velocity pattern and increase kinematic asymmetry.

In our analysis, we restrict the sample to galaxies classified as star-forming based on central optical emission-line diagnostics, thereby excluding systems with clear AGN signatures. Although weak or past AGN activity cannot be entirely ruled out, the dominant outflow contribution in our sample is expected to arise from starburst-driven winds.

Within this framework, a natural expectation is that galaxies with higher star formation rates should also exhibit stronger velocity map asymmetries, since more intense star formation drives more powerful winds. This prediction is consistent with our results, where galaxies above the star-forming main sequence show systematically enhanced kinematics asymmetry. By contrast, the correlation between kinematic asymmetry and gas-phase metallicity is unlikely to be explained by outflows. Stellar feedback both drives winds and enriches the interstellar medium, so galaxies experiencing strong starburst-driven outflows would be expected to have higher metallicities rather than the lower values we observe \citep{Chisholm2018, Langan2023}. Therefore, we speculate that outflows may dominate the velocity map asymmetry in a subset of galaxies, specifically those with SFRs significantly above the SFMS, metallicities not lower than the MZR, and relatively low $\hi$ fractions, as illustrated in Figure~\ref{fig:fmr_hi}.

It should be noted that, because galactic outflows are typically launched perpendicular to the disk \citep[e.g.,][]{Bouche2012}, their signatures are more easily detected in highly inclined, edge-on galaxies. Since our sample is dominated by low-inclination systems, the direct kinematic imprint of outflowing gas is likely to remain undetected unless the winds are especially strong. Nevertheless, if outflows inject turbulence on kiloparsec scales, they could still make a significant contribution to the measured velocity map asymmetry.

\subsubsection{Bar- and Spiral-driven Gas Streaming}

Non-axisymmetric structures such as stellar bars and spiral arms perturb the gravitational potential of disk galaxies and induce gas motions that deviate from pure circular rotation. The resulting streaming motion can be described by higher-order terms in harmonic decomposition, thereby contributing to enhanced kinematic asymmetry \citep{Schoenmakers1997, Wong2004, Spekkens2007}. In our sample, galaxies with prominent bars have been excluded based on photometric images, but a small fraction of galaxies with weaker bars may still remain. Thus, the influence of bars cannot be entirely ruled out, and perturbations driven by both bars and spiral arms are expected to leave imprints on the morphology of velocity maps.

Numerous studies have shown that bars and spiral arms can funnel gas from the outer disk toward the central regions, thereby enhancing central star formation rates \citep{Regan2004, Kim2014}. Observationally, galaxies with strong bars or prominent spiral structures are often found to exhibit more intense star formation activity \citep{Lin2017, Yu2021}. This bar- or spiral-driven gas streaming could therefore contribute to the observed link between elevated kinematic asymmetry and enhanced star formation. However, previous observations do not generally find that bars or spirals reduce gas-phase metallicity \citep{Ellison2008, SanchezMenguiano2016}. We therefore infer that the contribution of bars and spirals to the observed non-circular motions is not dominant.

\subsubsection{Galaxy Interactions and Mergers}

Tidal interactions and mergers are among the most powerful drivers of non-circular gas motions in galaxies. Gravitational torques generated during close encounters place gas on strongly perturbed orbits, producing pronounced asymmetries in the velocity map \citep[e.g.,][]{Hung2016, Feng2020}. Even after the final coalescence, such kinematic distortions can persist for gigayear timescales, leaving long-lived signatures that outlast obvious morphological disturbances \citep[e.g.,][]{Shapiro2008, McElroy2022}.  

Interactions and mergers also have a major impact on star formation and chemical enrichment. Tidal torques can drive strong gas inflows into the central regions, where the gas is compressed and fuels intense starbursts \citep[e.g.,][]{Li2008, Patton2013, Feng2024}. At the same time, companions or the intergalactic medium may provide relatively metal-poor material, diluting the gas-phase metallicity and producing offsets below the mass-metallicity relation \citep{MichelDansac2008, Scudder2012}. As a result, interactions and mergers can simultaneously trigger enhanced star formation, reduced metallicities, and strongly asymmetric velocity maps. 

In our sample, we exclude interacting pairs and clear merger remnants based on morphological and environmental criteria. Nevertheless, some post-merger systems and minor mergers below our detection thresholds may remain, indicating that their impact, while modest, could still play a non-negligible role in shaping the observed trends.

\subsubsection{Disk Warps and Local Turbulence}

It is important to note that not all asymmetries in velocity map morphology arise from classical non-circular motions such as inflows and outflows. Additional effects, including disk warps and small-scale turbulence, can also contribute to elevated kinematic asymmetry, but their influence is subdominant and unlikely to drive the global trends reported here.  

Disk warps are large-scale three-dimensional distortions in which the orientation of the disk mid-plane varies with radius. In projection, they appear as radial twists in photometric or kinematic position angles, often accompanied by changes in ellipticity. Such warps can contribute to kinematic asymmetry when the position angle and ellipticity are assumed to remain constant with radius \citep[e.g.,][]{Krajnovic2006, Wang2023}. However, they are typically most pronounced in the outer disk \citep{GarciaRuiz2002}, well beyond the effective radius, and their impact on the velocity map within $R_e$, the region relevant to our measurements, is expected to be negligible.

Local turbulence refers to random or chaotic gas motions on sub-kiloparsec scales, often driven by stellar feedback or gravitational instabilities. Given that the spatial resolution of MaNGA velocity maps is on the order of $\sim$1 kpc, turbulent structures on smaller scales remain unresolved. Their main observational imprint is therefore line broadening and elevated velocity dispersion, rather than coherent distortions of the velocity map. The turbulence on kiloparsec scales can indeed enhance the measured kinematic asymmetry, but such effects are likely to occur in conjunction with other mechanisms discussed above.

\section{Summary}\label{sec:sum}

In this paper, we investigated the morphology of ionized-gas velocity maps in a large sample of $528$ low-inclination, regular disk galaxies from the SDSS-IV MaNGA survey, with the goal of understanding how deviations from ordered disk rotation connect to global galaxy properties. Non-circular motions were quantified using harmonic decomposition, employing two complementary diagnostics: the global kinematic asymmetry $\vasym$, which traces non-axisymmetric perturbations, and the first-order term ratio $\vab$, which reflects axisymmetric radial motions.  

Our main observational results can be summarized as follows:  
\begin{itemize}
    \item The kinematic asymmetry parameter $\vasym$ shows strong correlations with both star formation rate (SFR) and gas-phase metallicity across the full stellar mass range. Galaxies with highly asymmetric velocity maps are preferentially displaced from the scaling relations, lying either above or below the star-forming main sequence (SFMS) and systematically below the mass-metallicity relation (MZR). Notably, the population of galaxies below the SFMS with elevated kinematic asymmetry appears only at the low-mass end ($\logm < 10.2$). An analysis of the $\vasym$ distributions in different regions of the SFMS and MZR planes, accounting for measurement uncertainties, confirms that $\vasym$ is significantly correlated with both SFR and metallicity.
    \item In contrast, the first-order term ratio $\vab$ exhibits only a limited connection with global galaxy properties: it shows a correlation with metallicity in the low-mass range ($\logm < 10.5$) but no significant dependence on SFR. After accounting for the relatively large measurement uncertainties, the $\vab$ distributions are statistically consistent with showing a relatively weak correlation with metallicity and no robust correlation with SFR. 
    \item Galaxies that fall below the MZR generally display elevated $\hi$ gas fractions and typically show higher kinematic asymmetry. However, this trend disappears among galaxies with strongly enhanced SFRs above the SFMS, where the $\hi$ fraction is suppressed while the velocity maps remain highly asymmetric.
\end{itemize}

To interpret these observational results, we explored the physical mechanisms capable of producing non-circular gas motions. We find that external gas accretion provides the most consistent explanation: it simultaneously dilutes metallicity, perturbs velocity maps, and supplies fresh fuel for star formation, thereby naturally linking velocity map asymmetry with SFR, metallicity, and $\hi$ fraction. Additional processes, including starburst-driven outflows, bar- and spiral-driven inflows, and interactions or mergers, can reproduce aspects of the observed trends and may contribute a part of the results, but they cannot account for the global correlations.  

\section*{Acknowlegements}
We thank the anonymous referee for valuable comments and suggestions that helped improve the quality of this paper. This work is supported by the National Natural Science Foundation of China (Nos. 12103017, 12073059, 12141302, 12333002, 12273051), Natural Science Foundation of Hebei Province (No. A2025205037), the project of the Hebei Provincial Department of Science and Technology (No. 226Z7604G), the National Key R\&D Program of China (Nos. 2022YFF0503402, 2019YFA0405501, 2022YFA1605300, 2022YFA1602901). SSY thanks research grants from China Manned Space Project with No. CMS-CSST-2021-A07, and Shanghai Academic/Technology Research Leader (22XD1404200). JY acknowledges support from the Natural Science Foundation of Shanghai (No. 22ZR1473000). 

Funding for the Sloan Digital Sky Survey IV has been provided by the Alfred P. Sloan Foundation, the U.S. Department of Energy Office of Science, and the Participating Institutions. SDSS-IV acknowledges support and resources from the Center for High-Performance Computing at the University of Utah. The SDSS website is \url{www.sdss.org}.

SDSS-IV is managed by the Astrophysical Research Consortium for the Participating Institutions of the SDSS Collaboration including the Brazilian Participation Group, the Carnegie Institution for Science, Carnegie Mellon University, the Chilean Participation Group, the French Participation Group, Harvard-Smithsonian Center for Astrophysics, Instituto de Astrof\'isica de Canarias, The Johns Hopkins University, Kavli Institute for the Physics and Mathematics of the Universe (IPMU) / University of Tokyo, the Korean Participation Group, Lawrence Berkeley National Laboratory, Leibniz Institut f\"ur Astrophysik Potsdam (AIP),  Max-Planck-Institut f\"ur Astronomie (MPIA Heidelberg), Max-Planck-Institut f\"ur Astrophysik (MPA Garching), Max-Planck-Institut f\"ur Extraterrestrische Physik (MPE), National Astronomical Observatories of China, New Mexico State University, New York University, University of Notre Dame, Observat\'ario Nacional / MCTI, The Ohio State University, Pennsylvania State University, Shanghai Astronomical Observatory, United Kingdom Participation Group, Universidad Nacional Aut\'onoma de M\'exico, University of Arizona, University of Colorado Boulder, University of Oxford, University of Portsmouth, University of Utah, University of Virginia, University of Washington, University of Wisconsin, Vanderbilt University, and Yale University.

\bibliography{ref}{}

\begin{thebibliography}{}
\expandafter\ifx\csname natexlab\endcsname\relax\def\natexlab#1{#1}\fi
\providecommand{\url}[1]{\href{#1}{#1}}
\providecommand{\dodoi}[1]{doi:~\href{http://doi.org/#1}{\nolinkurl{#1}}}
\providecommand{\doeprint}[1]{\href{http://ascl.net/#1}{\nolinkurl{http://ascl.net/#1}}}
\providecommand{\doarXiv}[1]{\href{https://arxiv.org/abs/#1}{\nolinkurl{https://arxiv.org/abs/#1}}}

\bibitem[{{Abdurro'uf} {et~al.}(2022){Abdurro'uf}, {Accetta}, {Aerts}, {Silva
  Aguirre}, {Ahumada}, {Ajgaonkar}, {Filiz Ak}, {Alam}, {Allende Prieto},
  {Almeida}, \& et~al.}]{SDSSDR17}
{Abdurro'uf}, {Accetta}, K., {Aerts}, C., {et~al.} 2022, \apjs, 259, 35,
  \dodoi{10.3847/1538-4365/ac4414}

\bibitem[{{Baker} {et~al.}(2023){Baker}, {Maiolino}, {Belfiore}, {Curti},
  {Bluck}, {Lin}, {Ellison}, {Thorp}, \& {Pan}}]{Baker2023}
{Baker}, W.~M., {Maiolino}, R., {Belfiore}, F., {et~al.} 2023, \mnras, 519,
  1149, \dodoi{10.1093/mnras/stac3594}

\bibitem[{{Belfiore} {et~al.}(2019){Belfiore}, {Westfall}, {Schaefer},
  {Cappellari}, {Ji}, {Bershady}, {Tremonti}, {Law}, {Yan}, {Bundy}, {Shetty},
  {Drory}, {Thomas}, {Emsellem}, \& {S{\'a}nchez}}]{Belfiore2019}
{Belfiore}, F., {Westfall}, K.~B., {Schaefer}, A., {et~al.} 2019, \aj, 158,
  160, \dodoi{10.3847/1538-3881/ab3e4e}

\bibitem[{{Blanton} {et~al.}(2017){Blanton}, {Bershady}, {Abolfathi},
  {Albareti}, {Allende Prieto}, {Almeida}, {Alonso-Garc{\'\i}a}, {Anders},
  {Anderson}, {Andrews}, {Aquino-Ort{\'\i}z}, {Arag{\'o}n-Salamanca},
  {Argudo-Fern{\'a}ndez}, {Armengaud}, {Aubourg}, {Avila-Reese}, {Badenes},
  {Bailey}, {Barger}, {Barrera-Ballesteros}, {Bartosz}, {Bates}, {Baumgarten},
  {Bautista}, {Beaton}, {Beers}, {Belfiore}, {Bender}, {Berlind}, {Bernardi},
  {Beutler}, {Bird}, {Bizyaev}, {Blanc}, {Blomqvist}, {Bolton}, {Boquien},
  {Borissova}, {van den Bosch}, {Bovy}, {Brandt}, {Brinkmann}, {Brownstein},
  {Bundy}, {Burgasser}, {Burtin}, {Busca}, {Cappellari}, {Delgado Carigi},
  {Carlberg}, {Carnero Rosell}, {Carrera}, {Chanover}, {Cherinka}, {Cheung},
  {G{\'o}mez Maqueo Chew}, {Chiappini}, {Choi}, {Chojnowski}, {Chuang},
  {Chung}, {Cirolini}, {Clerc}, {Cohen}, {Comparat}, {da Costa}, {Cousinou},
  {Covey}, {Crane}, {Croft}, {Cruz-Gonzalez}, {Garrido Cuadra}, {Cunha},
  {Damke}, {Darling}, {Davies}, {Dawson}, {de la Macorra}, {Dell'Agli}, {De
  Lee}, {Delubac}, {Di Mille}, {Diamond-Stanic}, {Cano-D{\'\i}az}, {Donor},
  {Downes}, {Drory}, {du Mas des Bourboux}, {Duckworth}, {Dwelly}, {Dyer},
  {Ebelke}, {Eigenbrot}, {Eisenstein}, {Emsellem}, {Eracleous}, {Escoffier},
  {Evans}, {Fan}, {Fern{\'a}ndez-Alvar}, {Fernandez-Trincado}, {Feuillet},
  {Finoguenov}, {Fleming}, {Font-Ribera}, {Fredrickson}, {Freischlad},
  {Frinchaboy}, {Fuentes}, {Galbany}, {Garcia-Dias},
  {Garc{\'\i}a-Hern{\'a}ndez}, {Gaulme}, {Geisler}, {Gelfand},
  {Gil-Mar{\'\i}n}, {Gillespie}, {Goddard}, {Gonzalez-Perez}, {Grabowski},
  {Green}, {Grier}, {Gunn}, {Guo}, {Guy}, {Hagen}, {Hahn}, {Hall}, {Harding},
  {Hasselquist}, {Hawley}, {Hearty}, {Gonzalez Hern{\'a}ndez}, {Ho}, {Hogg},
  {Holley-Bockelmann}, {Holtzman}, {Holzer}, {Huehnerhoff}, {Hutchinson},
  {Hwang}, {Ibarra-Medel}, {da Silva Ilha}, {Ivans}, {Ivory}, {Jackson},
  {Jensen}, {Johnson}, {Jones}, {J{\"o}nsson}, {Jullo}, {Kamble}, {Kinemuchi},
  {Kirkby}, {Kitaura}, {Klaene}, {Knapp}, {Kneib}, {Kollmeier}, {Lacerna},
  {Lane}, {Lang}, {Law}, {Lazarz}, {Lee}, {Le Goff}, {Liang}, {Li}, {Li},
  {Lian}, {Lima}, {Lin}, {Lin}, {Bertran de Lis}, {Liu}, {de Icaza Lizaola},
  {Long}, {Lucatello}, {Lundgren}, {MacDonald}, {Deconto Machado}, {MacLeod},
  {Mahadevan}, {Geimba Maia}, {Maiolino}, {Majewski}, {Malanushenko},
  {Malanushenko}, {Manchado}, {Mao}, {Maraston}, {Marques-Chaves}, {Masseron},
  {Masters}, {McBride}, {McDermid}, {McGrath}, {McGreer}, {Medina Pe{\~n}a},
  {Melendez}, {Merloni}, {Merrifield}, {Meszaros}, {Meza}, {Minchev},
  {Minniti}, {Miyaji}, {More}, {Mulchaey}, {M{\"u}ller-S{\'a}nchez}, {Muna},
  {Munoz}, {Myers}, {Nair}, {Nandra}, {Correa do Nascimento}, {Negrete},
  {Ness}, {Newman}, {Nichol}, {Nidever}, {Nitschelm}, {Ntelis}, {O'Connell},
  {Oelkers}, {Oravetz}, {Oravetz}, {Pace}, {Padilla}, {Palanque-Delabrouille},
  {Alonso Palicio}, {Pan}, {Parejko}, {Parikh}, {P{\^a}ris}, {Park}, {Patten},
  {Peirani}, {Pellejero-Ibanez}, {Penny}, {Percival}, {Perez-Fournon},
  {Petitjean}, {Pieri}, {Pinsonneault}, {Pisani}, {Poleski}, {Prada},
  {Prakash}, {Queiroz}, {Raddick}, {Raichoor}, {Barboza Rembold}, {Richstein},
  {Riffel}, {Riffel}, {Rix}, {Robin}, {Rockosi}, {Rodr{\'\i}guez-Torres},
  {Roman-Lopes}, {Rom{\'a}n-Z{\'u}{\~n}iga}, {Rosado}, {Ross}, {Rossi}, {Ruan},
  {Ruggeri}, {Rykoff}, {Salazar-Albornoz}, {Salvato}, {S{\'a}nchez}, {Aguado},
  {S{\'a}nchez-Gallego}, {Santana}, {Santiago}, {Sayres}, {Schiavon}, {da Silva
  Schimoia}, {Schlafly}, {Schlegel}, {Schneider}, {Schultheis}, {Schuster},
  {Schwope}, {Seo}, {Shao}, {Shen}, {Shetrone}, {Shull}, {Simon}, {Skinner},
  {Skrutskie}, {Slosar}, {Smith}, {Sobeck}, {Sobreira}, {Somers}, {Souto},
  {Stark}, {Stassun}, {Stauffer}, {Steinmetz}, {Storchi-Bergmann},
  {Streblyanska}, {Stringfellow}, {Su{\'a}rez}, {Sun}, {Suzuki}, {Szigeti},
  {Taghizadeh-Popp}, {Tang}, {Tao}, {Tayar}, {Tembe}, {Teske}, {Thakar},
  {Thomas}, {Thompson}, {Tinker}, {Tissera}, {Tojeiro}, {Hernandez Toledo}, {de
  la Torre}, {Tremonti}, {Troup}, {Valenzuela}, {Martinez Valpuesta},
  {Vargas-Gonz{\'a}lez}, {Vargas-Maga{\~n}a}, {Vazquez}, {Villanova}, {Vivek},
  {Vogt}, {Wake}, {Walterbos}, {Wang}, {Weaver}, {Weijmans}, {Weinberg},
  {Westfall}, {Whelan}, {Wild}, {Wilson}, {Wood-Vasey}, {Wylezalek}, {Xiao},
  {Yan}, {Yang}, {Ybarra}, {Y{\`e}che}, {Zakamska}, {Zamora}, {Zarrouk},
  {Zasowski}, {Zhang}, {Zhao}, {Zheng}, {Zheng}, {Zhou}, {Zhou}, {Zhu},
  {Zoccali}, \& {Zou}}]{Blanton2017}
{Blanton}, M.~R., {Bershady}, M.~A., {Abolfathi}, B., {et~al.} 2017, \aj, 154,
  28, \dodoi{10.3847/1538-3881/aa7567}

\bibitem[{{Bloom} {et~al.}(2017){Bloom}, {Fogarty}, {Croom}, {Schaefer},
  {Bryant}, {Cortese}, {Richards}, {Bland-Hawthorn}, {Ho}, {Scott},
  {Goldstein}, {Medling}, {Brough}, {Sweet}, {Cecil}, {L{\'o}pez-S{\'a}nchez},
  {Glazebrook}, {Parker}, {Allen}, {Goodwin}, {Green}, {Konstantopoulos},
  {Lawrence}, {Lorente}, {Owers}, \& {Sharp}}]{Bloom2017}
{Bloom}, J.~V., {Fogarty}, L.~M.~R., {Croom}, S.~M., {et~al.} 2017, \mnras,
  465, 123, \dodoi{10.1093/mnras/stw2605}

\bibitem[{{Bloom} {et~al.}(2018){Bloom}, {Croom}, {Bryant}, {Schaefer},
  {Bland-Hawthorn}, {Brough}, {Callingham}, {Cortese}, {Federrath}, {Scott},
  {van de Sande}, {D'Eugenio}, {Sweet}, {Tonini}, {Allen}, {Goodwin}, {Green},
  {Konstantopoulos}, {Lawrence}, {Lorente}, {Medling}, {Owers}, {Richards}, \&
  {Sharp}}]{Bloom2018}
{Bloom}, J.~V., {Croom}, S.~M., {Bryant}, J.~J., {et~al.} 2018, \mnras, 476,
  2339, \dodoi{10.1093/mnras/sty273}

\bibitem[{{Bouch{\'e}} {et~al.}(2012){Bouch{\'e}}, {Hohensee}, {Vargas},
  {Kacprzak}, {Martin}, {Cooke}, \& {Churchill}}]{Bouche2012}
{Bouch{\'e}}, N., {Hohensee}, W., {Vargas}, R., {et~al.} 2012, \mnras, 426,
  801, \dodoi{10.1111/j.1365-2966.2012.21114.x}

\bibitem[{{Brinchmann} {et~al.}(2004){Brinchmann}, {Charlot}, {White},
  {Tremonti}, {Kauffmann}, {Heckman}, \& {Brinkmann}}]{Brinchmann2004}
{Brinchmann}, J., {Charlot}, S., {White}, S.~D.~M., {et~al.} 2004, \mnras, 351,
  1151, \dodoi{10.1111/j.1365-2966.2004.07881.x}

\bibitem[{{Bryant} {et~al.}(2015){Bryant}, {Owers}, {Robotham}, {Croom},
  {Driver}, {Drinkwater}, {Lorente}, {Cortese}, {Scott}, {Colless}, {Schaefer},
  {Taylor}, {Konstantopoulos}, {Allen}, {Baldry}, {Barnes}, {Bauer},
  {Bland-Hawthorn}, {Bloom}, {Brooks}, {Brough}, {Cecil}, {Couch}, {Croton},
  {Davies}, {Ellis}, {Fogarty}, {Foster}, {Glazebrook}, {Goodwin}, {Green},
  {Gunawardhana}, {Hampton}, {Ho}, {Hopkins}, {Kewley}, {Lawrence},
  {Leon-Saval}, {Leslie}, {McElroy}, {Lewis}, {Liske}, {L{\'o}pez-S{\'a}nchez},
  {Mahajan}, {Medling}, {Metcalfe}, {Meyer}, {Mould}, {Obreschkow}, {O'Toole},
  {Pracy}, {Richards}, {Shanks}, {Sharp}, {Sweet}, {Thomas}, {Tonini}, \&
  {Walcher}}]{Bryant2015}
{Bryant}, J.~J., {Owers}, M.~S., {Robotham}, A.~S.~G., {et~al.} 2015, \mnras,
  447, 2857, \dodoi{10.1093/mnras/stu2635}

\bibitem[{{Bundy} {et~al.}(2015){Bundy}, {Bershady}, {Law}, {Yan}, {Drory},
  {MacDonald}, {Wake}, {Cherinka}, {S{\'a}nchez-Gallego}, {Weijmans}, {Thomas},
  {Tremonti}, {Masters}, {Coccato}, {Diamond-Stanic}, {Arag{\'o}n-Salamanca},
  {Avila-Reese}, {Badenes}, {Falc{\'o}n-Barroso}, {Belfiore}, {Bizyaev},
  {Blanc}, {Bland-Hawthorn}, {Blanton}, {Brownstein}, {Byler}, {Cappellari},
  {Conroy}, {Dutton}, {Emsellem}, {Etherington}, {Frinchaboy}, {Fu}, {Gunn},
  {Harding}, {Johnston}, {Kauffmann}, {Kinemuchi}, {Klaene}, {Knapen},
  {Leauthaud}, {Li}, {Lin}, {Maiolino}, {Malanushenko}, {Malanushenko}, {Mao},
  {Maraston}, {McDermid}, {Merrifield}, {Nichol}, {Oravetz}, {Pan}, {Parejko},
  {Sanchez}, {Schlegel}, {Simmons}, {Steele}, {Steinmetz}, {Thanjavur},
  {Thompson}, {Tinker}, {van den Bosch}, {Westfall}, {Wilkinson}, {Wright},
  {Xiao}, \& {Zhang}}]{Bundy2015}
{Bundy}, K., {Bershady}, M.~A., {Law}, D.~R., {et~al.} 2015, \apj, 798, 7,
  \dodoi{10.1088/0004-637X/798/1/7}

\bibitem[{{Cappellari} {et~al.}(2013){Cappellari}, {McDermid}, {Alatalo},
  {Blitz}, {Bois}, {Bournaud}, {Bureau}, {Crocker}, {Davies}, {Davis}, {de
  Zeeuw}, {Duc}, {Emsellem}, {Khochfar}, {Krajnovi{\'c}}, {Kuntschner},
  {Morganti}, {Naab}, {Oosterloo}, {Sarzi}, {Scott}, {Serra}, {Weijmans}, \&
  {Young}}]{Cappellari2013}
{Cappellari}, M., {McDermid}, R.~M., {Alatalo}, K., {et~al.} 2013, MNRAS, 432,
  1862, \dodoi{10.1093/mnras/stt644}

\bibitem[{{Chisholm} {et~al.}(2018){Chisholm}, {Tremonti}, \&
  {Leitherer}}]{Chisholm2018}
{Chisholm}, J., {Tremonti}, C., \& {Leitherer}, C. 2018, \mnras, 481, 1690,
  \dodoi{10.1093/mnras/sty2380}

\bibitem[{Cleveland(1979)}]{Cleveland1979}
Cleveland, W.~S. 1979, Journal of the American Statistical Association, 74,
  829, \dodoi{10.1080/01621459.1979.10481038}

\bibitem[{{Cresci} {et~al.}(2019){Cresci}, {Mannucci}, \& {Curti}}]{Cresci2019}
{Cresci}, G., {Mannucci}, F., \& {Curti}, M. 2019, \aap, 627, A42,
  \dodoi{10.1051/0004-6361/201834637}

\bibitem[{{Danovich} {et~al.}(2015){Danovich}, {Dekel}, {Hahn}, {Ceverino}, \&
  {Primack}}]{Danovich2015}
{Danovich}, M., {Dekel}, A., {Hahn}, O., {Ceverino}, D., \& {Primack}, J. 2015,
  \mnras, 449, 2087, \dodoi{10.1093/mnras/stv270}

\bibitem[{{Dav{\'e}} {et~al.}(2011){Dav{\'e}}, {Oppenheimer}, \&
  {Finlator}}]{Dave2011}
{Dav{\'e}}, R., {Oppenheimer}, B.~D., \& {Finlator}, K. 2011, \mnras, 415, 11,
  \dodoi{10.1111/j.1365-2966.2011.18680.x}

\bibitem[{{Dekel} \& {Birnboim}(2006)}]{Dekel2006}
{Dekel}, A., \& {Birnboim}, Y. 2006, \mnras, 368, 2,
  \dodoi{10.1111/j.1365-2966.2006.10145.x}

\bibitem[{{Di Teodoro} \& {Fraternali}(2015)}]{DiTeodoro2015}
{Di Teodoro}, E.~M., \& {Fraternali}, F. 2015, \mnras, 451, 3021,
  \dodoi{10.1093/mnras/stv1213}

\bibitem[{{Di Teodoro} \& {Peek}(2021)}]{DiTeodoro2021}
{Di Teodoro}, E.~M., \& {Peek}, J.~E.~G. 2021, \apj, 923, 220,
  \dodoi{10.3847/1538-4357/ac2cbd}

\bibitem[{{Drory} {et~al.}(2015){Drory}, {MacDonald}, {Bershady}, {Bundy},
  {Gunn}, {Law}, {Smith}, {Stoll}, {Tremonti}, {Wake}, {Yan}, {Weijmans},
  {Byler}, {Cherinka}, {Cope}, {Eigenbrot}, {Harding}, {Holder}, {Huehnerhoff},
  {Jaehnig}, {Jansen}, {Klaene}, {Paat}, {Percival}, \& {Sayres}}]{Drory2015}
{Drory}, N., {MacDonald}, N., {Bershady}, M.~A., {et~al.} 2015, \aj, 149, 77,
  \dodoi{10.1088/0004-6256/149/2/77}

\bibitem[{{Ellison} {et~al.}(2008){Ellison}, {Patton}, {Simard}, \&
  {McConnachie}}]{Ellison2008}
{Ellison}, S.~L., {Patton}, D.~R., {Simard}, L., \& {McConnachie}, A.~W. 2008,
  \apjl, 672, L107, \dodoi{10.1086/527296}

\bibitem[{{Fathi} {et~al.}(2006){Fathi}, {Storchi-Bergmann}, {Riffel}, {Winge},
  {Axon}, {Robinson}, {Capetti}, \& {Marconi}}]{Fathi2006}
{Fathi}, K., {Storchi-Bergmann}, T., {Riffel}, R.~A., {et~al.} 2006, \apjl,
  641, L25, \dodoi{10.1086/503832}

\bibitem[{{Feng} {et~al.}(2022){Feng}, {Shen}, {Yuan}, {Dai}, \&
  {Masters}}]{Feng2022}
{Feng}, S., {Shen}, S.-Y., {Yuan}, F.-T., {Dai}, Y.~S., \& {Masters}, K.~L.
  2022, \apjs, 262, 6, \dodoi{10.3847/1538-4365/ac80f2}

\bibitem[{{Feng} {et~al.}(2020){Feng}, {Shen}, {Yuan}, {Riffel}, \&
  {Pan}}]{Feng2020}
{Feng}, S., {Shen}, S.-Y., {Yuan}, F.-T., {Riffel}, R.~A., \& {Pan}, K. 2020,
  \apjl, 892, L20, \dodoi{10.3847/2041-8213/ab7dba}

\bibitem[{{Feng} {et~al.}(2024){Feng}, {Shen}, {Yuan}, {Zhong}, {Cui}, \&
  {Li}}]{Feng2024}
{Feng}, S., {Shen}, S.-Y., {Yuan}, F.-T., {et~al.} 2024, \apj, 965, 60,
  \dodoi{10.3847/1538-4357/ad343e}

\bibitem[{{Feng} {et~al.}(2019){Feng}, {Shen}, {Yuan}, {Luo}, {Zhang}, {Wang},
  {Wang}, {Li}, {Hou}, {Kong}, {Guo}, \& {Zuo}}]{Feng2019}
---. 2019, \apj, 880, 114, \dodoi{10.3847/1538-4357/ab24da}

\bibitem[{{Finlator} \& {Dav{\'e}}(2008)}]{Finlator2008}
{Finlator}, K., \& {Dav{\'e}}, R. 2008, \mnras, 385, 2181,
  \dodoi{10.1111/j.1365-2966.2008.12991.x}

\bibitem[{{Flores} {et~al.}(2006){Flores}, {Hammer}, {Puech}, {Amram}, \&
  {Balkowski}}]{Flores2006}
{Flores}, H., {Hammer}, F., {Puech}, M., {Amram}, P., \& {Balkowski}, C. 2006,
  \aap, 455, 107, \dodoi{10.1051/0004-6361:20054217}

\bibitem[{{F{\"o}rster Schreiber} \& {Wuyts}(2020)}]{ForsterSchreiber2020}
{F{\"o}rster Schreiber}, N.~M., \& {Wuyts}, S. 2020, \araa, 58, 661,
  \dodoi{10.1146/annurev-astro-032620-021910}

\bibitem[{{F{\"o}rster Schreiber} {et~al.}(2009){F{\"o}rster Schreiber},
  {Genzel}, {Bouch{\'e}}, {Cresci}, {Davies}, {Buschkamp}, {Shapiro},
  {Tacconi}, {Hicks}, {Genel}, {Shapley}, {Erb}, {Steidel}, {Lutz},
  {Eisenhauer}, {Gillessen}, {Sternberg}, {Renzini}, {Cimatti}, {Daddi},
  {Kurk}, {Lilly}, {Kong}, {Lehnert}, {Nesvadba}, {Verma}, {McCracken},
  {Arimoto}, {Mignoli}, \& {Onodera}}]{ForsterSchreiber2009}
{F{\"o}rster Schreiber}, N.~M., {Genzel}, R., {Bouch{\'e}}, N., {et~al.} 2009,
  \apj, 706, 1364, \dodoi{10.1088/0004-637X/706/2/1364}

\bibitem[{{Franx} {et~al.}(1994){Franx}, {van Gorkom}, \& {de
  Zeeuw}}]{Franx1994}
{Franx}, M., {van Gorkom}, J.~H., \& {de Zeeuw}, T. 1994, \apj, 436, 642,
  \dodoi{10.1086/174939}

\bibitem[{{Garc{\'\i}a-Lorenzo} {et~al.}(2015){Garc{\'\i}a-Lorenzo},
  {M{\'a}rquez}, {Barrera-Ballesteros}, {Masegosa}, {Husemann},
  {Falc{\'o}n-Barroso}, {Lyubenova}, {S{\'a}nchez}, {Walcher}, {Mast},
  {Garc{\'\i}a-Benito}, {M{\'e}ndez-Abreu}, {van de Ven}, {Spekkens}, {Holmes},
  {Monreal-Ibero}, {del Olmo}, {Ziegler}, {Bland-Hawthorn},
  {S{\'a}nchez-Bl{\'a}zquez}, {Iglesias-P{\'a}ramo}, {Aguerri}, {Papaderos},
  {Gomes}, {Marino}, {Gonz{\'a}lez Delgado}, {Cortijo-Ferrero},
  {L{\'o}pez-S{\'a}nchez}, {Bekerait{\.{e}}}, {Wisotzki}, \&
  {Bomans}}]{GarciaLorenzo2015}
{Garc{\'\i}a-Lorenzo}, B., {M{\'a}rquez}, I., {Barrera-Ballesteros}, J.~K.,
  {et~al.} 2015, \aap, 573, A59, \dodoi{10.1051/0004-6361/201423485}

\bibitem[{{Garc{\'\i}a-Ruiz} {et~al.}(2002){Garc{\'\i}a-Ruiz}, {Sancisi}, \&
  {Kuijken}}]{GarciaRuiz2002}
{Garc{\'\i}a-Ruiz}, I., {Sancisi}, R., \& {Kuijken}, K. 2002, \aap, 394, 769,
  \dodoi{10.1051/0004-6361:20020976}

\bibitem[{{Genzel} {et~al.}(2011){Genzel}, {Newman}, {Jones}, {F{\"o}rster
  Schreiber}, {Shapiro}, {Genel}, {Lilly}, {Renzini}, {Tacconi}, {Bouch{\'e}},
  {Burkert}, {Cresci}, {Buschkamp}, {Carollo}, {Ceverino}, {Davies}, {Dekel},
  {Eisenhauer}, {Hicks}, {Kurk}, {Lutz}, {Mancini}, {Naab}, {Peng},
  {Sternberg}, {Vergani}, \& {Zamorani}}]{Genzel2011}
{Genzel}, R., {Newman}, S., {Jones}, T., {et~al.} 2011, \apj, 733, 101,
  \dodoi{10.1088/0004-637X/733/2/101}

\bibitem[{{Genzel} {et~al.}(2023){Genzel}, {Jolly}, {Liu}, {Price}, {Lee},
  {F{\"o}rster Schreiber}, {Tacconi}, {Herrera-Camus}, {Barfety}, {Burkert},
  {Cao}, {Davies}, {Dekel}, {Lee}, {Lutz}, {Naab}, {Neri}, {Nestor Shachar},
  {Pastras}, {Pulsoni}, {Renzini}, {Schuster}, {Shimizu}, {Stanley},
  {Sternberg}, \& {{\"U}bler}}]{Genzel2023}
{Genzel}, R., {Jolly}, J.~B., {Liu}, D., {et~al.} 2023, \apj, 957, 48,
  \dodoi{10.3847/1538-4357/acef1a}

\bibitem[{{Glazebrook}(2013)}]{Glazebrook2013}
{Glazebrook}, K. 2013, \pasa, 30, e056, \dodoi{10.1017/pasa.2013.34}

\bibitem[{{Gunn} {et~al.}(2006){Gunn}, {Siegmund}, {Mannery}, {Owen}, {Hull},
  {Leger}, {Carey}, {Knapp}, {York}, {Boroski}, {Kent}, {Lupton}, {Rockosi},
  {Evans}, {Waddell}, {Anderson}, {Annis}, {Barentine}, {Bartoszek}, {Bastian},
  {Bracker}, {Brewington}, {Briegel}, {Brinkmann}, {Brown}, {Carr},
  {Czarapata}, {Drennan}, {Dombeck}, {Federwitz}, {Gillespie}, {Gonzales},
  {Hansen}, {Harvanek}, {Hayes}, {Jordan}, {Kinney}, {Klaene}, {Kleinman},
  {Kron}, {Kresinski}, {Lee}, {Limmongkol}, {Lindenmeyer}, {Long}, {Loomis},
  {McGehee}, {Mantsch}, {Neilsen}, {Neswold}, {Newman}, {Nitta}, {Peoples},
  {Pier}, {Prieto}, {Prosapio}, {Rivetta}, {Schneider}, {Snedden}, \&
  {Wang}}]{Gunn2006}
{Gunn}, J.~E., {Siegmund}, W.~A., {Mannery}, E.~J., {et~al.} 2006, \aj, 131,
  2332, \dodoi{10.1086/500975}

\bibitem[{{Ho} {et~al.}(2014){Ho}, {Kewley}, {Dopita}, {Medling}, {Allen},
  {Bland-Hawthorn}, {Bloom}, {Bryant}, {Croom}, {Fogarty}, {Goodwin}, {Green},
  {Konstantopoulos}, {Lawrence}, {L{\'o}pez-S{\'a}nchez}, {Owers}, {Richards},
  \& {Sharp}}]{Ho2014}
{Ho}, I.~T., {Kewley}, L.~J., {Dopita}, M.~A., {et~al.} 2014, \mnras, 444,
  3894, \dodoi{10.1093/mnras/stu1653}

\bibitem[{{Ho} {et~al.}(2016){Ho}, {Medling}, {Bland-Hawthorn}, {Groves},
  {Kewley}, {Kobayashi}, {Dopita}, {Leslie}, {Sharp}, {Allen}, {Bourne},
  {Bryant}, {Cortese}, {Croom}, {Dunne}, {Fogarty}, {Goodwin}, {Green},
  {Konstantopoulos}, {Lawrence}, {Lorente}, {Owers}, {Richards}, {Sweet},
  {Tescari}, \& {Valiante}}]{Ho2016}
{Ho}, I.~T., {Medling}, A.~M., {Bland-Hawthorn}, J., {et~al.} 2016, \mnras,
  457, 1257, \dodoi{10.1093/mnras/stw017}

\bibitem[{{Hopkins} {et~al.}(2012){Hopkins}, {Quataert}, \&
  {Murray}}]{Hopkins2012}
{Hopkins}, P.~F., {Quataert}, E., \& {Murray}, N. 2012, \mnras, 421, 3522,
  \dodoi{10.1111/j.1365-2966.2012.20593.x}

\bibitem[{{Hung} {et~al.}(2016){Hung}, {Hayward}, {Smith}, {Ashby}, {Lanz},
  {Mart{\'\i}nez-Galarza}, {Sanders}, \& {Zezas}}]{Hung2016}
{Hung}, C.-L., {Hayward}, C.~C., {Smith}, H.~A., {et~al.} 2016, \apj, 816, 99,
  \dodoi{10.3847/0004-637X/816/2/99}

\bibitem[{{Jin} {et~al.}(2016){Jin}, {Chen}, {Shi}, {Tremonti}, {Bershady},
  {Merrifield}, {Emsellem}, {Fu}, {Wake}, {Bundy}, {Lin}, {Argudo-Fernandez},
  {Huang}, {Stark}, {Storchi-Bergmann}, {Bizyaev}, {Brownstein}, {Chisholm},
  {Guo}, {Hao}, {Hu}, {Li}, {Li}, {Masters}, {Malanushenko}, {Pan}, {Riffel},
  {Roman-Lopes}, {Simmons}, {Thomas}, {Wang}, {Westfall}, \& {Yan}}]{Jin2016}
{Jin}, Y., {Chen}, Y., {Shi}, Y., {et~al.} 2016, \mnras, 463, 913,
  \dodoi{10.1093/mnras/stw2055}

\bibitem[{{Kauffmann} {et~al.}(2003{\natexlab{a}}){Kauffmann}, {Heckman},
  {White}, {Charlot}, {Tremonti}, {Brinchmann}, {Bruzual}, {Peng}, {Seibert},
  {Bernardi}, {Blanton}, {Brinkmann}, {Castander}, {Cs{\'a}bai}, {Fukugita},
  {Ivezic}, {Munn}, {Nichol}, {Padmanabhan}, {Thakar}, {Weinberg}, \&
  {York}}]{Kauffmann2003b}
{Kauffmann}, G., {Heckman}, T.~M., {White}, S. D.~M., {et~al.}
  2003{\natexlab{a}}, \mnras, 341, 33, \dodoi{10.1046/j.1365-8711.2003.06291.x}

\bibitem[{{Kauffmann} {et~al.}(2003{\natexlab{b}}){Kauffmann}, {Heckman},
  {Tremonti}, {Brinchmann}, {Charlot}, {White}, {Ridgway}, {Brinkmann},
  {Fukugita}, {Hall}, {Ivezi{\'c}}, {Richards}, \&
  {Schneider}}]{Kauffmann2003a}
{Kauffmann}, G., {Heckman}, T.~M., {Tremonti}, C., {et~al.} 2003{\natexlab{b}},
  \mnras, 346, 1055, \dodoi{10.1111/j.1365-2966.2003.07154.x}

\bibitem[{{Kere{\v{s}}} {et~al.}(2005){Kere{\v{s}}}, {Katz}, {Weinberg}, \&
  {Dav{\'e}}}]{Keres2005}
{Kere{\v{s}}}, D., {Katz}, N., {Weinberg}, D.~H., \& {Dav{\'e}}, R. 2005,
  \mnras, 363, 2, \dodoi{10.1111/j.1365-2966.2005.09451.x}

\bibitem[{{Kewley} {et~al.}(2006){Kewley}, {Groves}, {Kauffmann}, \&
  {Heckman}}]{Kewley2006}
{Kewley}, L.~J., {Groves}, B., {Kauffmann}, G., \& {Heckman}, T. 2006, \mnras,
  372, 961, \dodoi{10.1111/j.1365-2966.2006.10859.x}

\bibitem[{{Khoram} \& {Belfiore}(2025)}]{Khoram2025}
{Khoram}, A.~H., \& {Belfiore}, F. 2025, \aap, 693, A150,
  \dodoi{10.1051/0004-6361/202451980}

\bibitem[{{Kim} \& {Kim}(2014)}]{Kim2014}
{Kim}, Y., \& {Kim}, W.-T. 2014, \mnras, 440, 208, \dodoi{10.1093/mnras/stu276}

\bibitem[{{Krajnovi{\'c}} {et~al.}(2006){Krajnovi{\'c}}, {Cappellari}, {de
  Zeeuw}, \& {Copin}}]{Krajnovic2006}
{Krajnovi{\'c}}, D., {Cappellari}, M., {de Zeeuw}, P.~T., \& {Copin}, Y. 2006,
  \mnras, 366, 787, \dodoi{10.1111/j.1365-2966.2005.09902.x}

\bibitem[{{Langan} {et~al.}(2023){Langan}, {Zabl}, {Bouch{\'e}}, {Ginolfi},
  {Popping}, {Schroetter}, {Wendt}, {Schaye}, {Boogaard}, {Freundlich},
  {Richard}, {Matthee}, {Mercier}, {Contini}, {Guo}, \& {Cherrey}}]{Langan2023}
{Langan}, I., {Zabl}, J., {Bouch{\'e}}, N.~F., {et~al.} 2023, \mnras, 521, 546,
  \dodoi{10.1093/mnras/stad357}

\bibitem[{{Law} {et~al.}(2009){Law}, {Steidel}, {Erb}, {Larkin}, {Pettini},
  {Shapley}, \& {Wright}}]{Law2009}
{Law}, D.~R., {Steidel}, C.~C., {Erb}, D.~K., {et~al.} 2009, \apj, 697, 2057,
  \dodoi{10.1088/0004-637X/697/2/2057}

\bibitem[{{Law} {et~al.}(2015){Law}, {Yan}, {Bershady}, {Bundy}, {Cherinka},
  {Drory}, {MacDonald}, {S{\'a}nchez-Gallego}, {Wake}, {Weijmans}, {Blanton},
  {Klaene}, {Moran}, {Sanchez}, \& {Zhang}}]{Law2015}
{Law}, D.~R., {Yan}, R., {Bershady}, M.~A., {et~al.} 2015, \aj, 150, 19,
  \dodoi{10.1088/0004-6256/150/1/19}

\bibitem[{{Li} {et~al.}(2008){Li}, {Kauffmann}, {Heckman}, {Jing}, \&
  {White}}]{Li2008}
{Li}, C., {Kauffmann}, G., {Heckman}, T.~M., {Jing}, Y.~P., \& {White}, S.
  D.~M. 2008, \mnras, 385, 1903, \dodoi{10.1111/j.1365-2966.2008.13000.x}

\bibitem[{{Lilly} {et~al.}(2013){Lilly}, {Carollo}, {Pipino}, {Renzini}, \&
  {Peng}}]{Lilly2013}
{Lilly}, S.~J., {Carollo}, C.~M., {Pipino}, A., {Renzini}, A., \& {Peng}, Y.
  2013, \apj, 772, 119, \dodoi{10.1088/0004-637X/772/2/119}

\bibitem[{{Lin} {et~al.}(2017){Lin}, {Li}, {He}, {Xiao}, \& {Wang}}]{Lin2017}
{Lin}, L., {Li}, C., {He}, Y., {Xiao}, T., \& {Wang}, E. 2017, \apj, 838, 105,
  \dodoi{10.3847/1538-4357/aa657a}

\bibitem[{{Mannucci} {et~al.}(2010){Mannucci}, {Cresci}, {Maiolino}, {Marconi},
  \& {Gnerucci}}]{Mannucci2010}
{Mannucci}, F., {Cresci}, G., {Maiolino}, R., {Marconi}, A., \& {Gnerucci}, A.
  2010, \mnras, 408, 2115, \dodoi{10.1111/j.1365-2966.2010.17291.x}

\bibitem[{{Martin} {et~al.}(2019){Martin}, {O'Sullivan}, {Matuszewski},
  {Hamden}, {Dekel}, {Lapiner}, {Morrissey}, {Neill}, {Cantalupo}, {Prochaska},
  {Steidel}, {Trainor}, {Moore}, {Ceverino}, {Primack}, \&
  {Rizzi}}]{Martin2019}
{Martin}, D.~C., {O'Sullivan}, D., {Matuszewski}, M., {et~al.} 2019, Nature
  Astronomy, 3, 822, \dodoi{10.1038/s41550-019-0791-2}

\bibitem[{{Masters} {et~al.}(2019){Masters}, {Stark}, {Pace}, {Phipps},
  {Rujopakarn}, {Samanso}, {Harrington}, {S{\'a}nchez-Gallego}, {Avila-Reese},
  {Bershady}, {Cherinka}, {Fielder}, {Finnegan}, {Riffel}, {Rowlands},
  {Shamsi}, {Newnham}, {Weijmans}, \& {Witherspoon}}]{Masters2019}
{Masters}, K.~L., {Stark}, D.~V., {Pace}, Z.~J., {et~al.} 2019, \mnras, 488,
  3396, \dodoi{10.1093/mnras/stz1889}

\bibitem[{{McElroy} {et~al.}(2022){McElroy}, {Bottrell}, {Hani}, {Moreno},
  {Croom}, {Hayward}, {Twum}, {Feldmann}, {Hopkins}, {Hernquist}, \&
  {Husemann}}]{McElroy2022}
{McElroy}, R., {Bottrell}, C., {Hani}, M.~H., {et~al.} 2022, \mnras, 515, 3406,
  \dodoi{10.1093/mnras/stac1715}

\bibitem[{{Michel-Dansac} {et~al.}(2008){Michel-Dansac}, {Lambas}, {Alonso}, \&
  {Tissera}}]{MichelDansac2008}
{Michel-Dansac}, L., {Lambas}, D.~G., {Alonso}, M.~S., \& {Tissera}, P. 2008,
  \mnras, 386, L82, \dodoi{10.1111/j.1745-3933.2008.00466.x}

\bibitem[{{M{\"u}ller-S{\'a}nchez} {et~al.}(2011){M{\"u}ller-S{\'a}nchez},
  {Prieto}, {Hicks}, {Vives-Arias}, {Davies}, {Malkan}, {Tacconi}, \&
  {Genzel}}]{MullerSanchez2011}
{M{\"u}ller-S{\'a}nchez}, F., {Prieto}, M.~A., {Hicks}, E.~K.~S., {et~al.}
  2011, \apj, 739, 69, \dodoi{10.1088/0004-637X/739/2/69}

\bibitem[{{Naab} \& {Ostriker}(2017)}]{Naab2017}
{Naab}, T., \& {Ostriker}, J.~P. 2017, \araa, 55, 59,
  \dodoi{10.1146/annurev-astro-081913-040019}

\bibitem[{{Patton} {et~al.}(2013){Patton}, {Torrey}, {Ellison}, {Mendel}, \&
  {Scudder}}]{Patton2013}
{Patton}, D.~R., {Torrey}, P., {Ellison}, S.~L., {Mendel}, J.~T., \& {Scudder},
  J.~M. 2013, \mnras, 433, L59, \dodoi{10.1093/mnrasl/slt058}

\bibitem[{{Pettini} \& {Pagel}(2004)}]{PP04}
{Pettini}, M., \& {Pagel}, B. E.~J. 2004, \mnras, 348, L59,
  \dodoi{10.1111/j.1365-2966.2004.07591.x}

\bibitem[{{Regan} \& {Teuben}(2004)}]{Regan2004}
{Regan}, M.~W., \& {Teuben}, P.~J. 2004, \apj, 600, 595, \dodoi{10.1086/380116}

\bibitem[{{Salim} {et~al.}(2007){Salim}, {Rich}, {Charlot}, {Brinchmann},
  {Johnson}, {Schiminovich}, {Seibert}, {Mallery}, {Heckman}, {Forster},
  {Friedman}, {Martin}, {Morrissey}, {Neff}, {Small}, {Wyder}, {Bianchi},
  {Donas}, {Lee}, {Madore}, {Milliard}, {Szalay}, {Welsh}, \& {Yi}}]{Salim2007}
{Salim}, S., {Rich}, R.~M., {Charlot}, S., {et~al.} 2007, \apjs, 173, 267,
  \dodoi{10.1086/519218}

\bibitem[{{S{\'a}nchez} {et~al.}(2012){S{\'a}nchez}, {Kennicutt}, {Gil de Paz},
  {van de Ven}, {V{\'\i}lchez}, {Wisotzki}, {Walcher}, {Mast}, {Aguerri},
  {Albiol-P{\'e}rez}, {Alonso-Herrero}, {Alves}, {Bakos}, {Bart{\'a}kov{\'a}},
  {Bland-Hawthorn}, {Boselli}, {Bomans}, {Castillo-Morales}, {Cortijo-Ferrero},
  {de Lorenzo-C{\'a}ceres}, {Del Olmo}, {Dettmar}, {D{\'\i}az}, {Ellis},
  {Falc{\'o}n-Barroso}, {Flores}, {Gallazzi}, {Garc{\'\i}a-Lorenzo},
  {Gonz{\'a}lez Delgado}, {Gruel}, {Haines}, {Hao}, {Husemann},
  {Igl{\'e}sias-P{\'a}ramo}, {Jahnke}, {Johnson}, {Jungwiert}, {Kalinova},
  {Kehrig}, {Kupko}, {L{\'o}pez-S{\'a}nchez}, {Lyubenova}, {Marino},
  {M{\'a}rmol-Queralt{\'o}}, {M{\'a}rquez}, {Masegosa}, {Meidt},
  {Mendez-Abreu}, {Monreal-Ibero}, {Montijo}, {Mour{\~a}o}, {Palacios-Navarro},
  {Papaderos}, {Pasquali}, {Peletier}, {P{\'e}rez}, {P{\'e}rez}, {Quirrenbach},
  {Rela{\~n}o}, {Rosales-Ortega}, {Roth}, {Ruiz-Lara},
  {S{\'a}nchez-Bl{\'a}zquez}, {Sengupta}, {Singh}, {Stanishev}, {Trager},
  {Vazdekis}, {Viironen}, {Wild}, {Zibetti}, \& {Ziegler}}]{Sanchez2012}
{S{\'a}nchez}, S.~F., {Kennicutt}, R.~C., {Gil de Paz}, A., {et~al.} 2012,
  \aap, 538, A8, \dodoi{10.1051/0004-6361/201117353}

\bibitem[{{S{\'a}nchez} {et~al.}(2013){S{\'a}nchez}, {Rosales-Ortega},
  {Jungwiert}, {Iglesias-P{\'a}ramo}, {V{\'\i}lchez}, {Marino}, {Walcher},
  {Husemann}, {Mast}, {Monreal-Ibero}, {Cid Fernandes}, {P{\'e}rez},
  {Gonz{\'a}lez Delgado}, {Garc{\'\i}a-Benito}, {Galbany}, {van de Ven},
  {Jahnke}, {Flores}, {Bland-Hawthorn}, {L{\'o}pez-S{\'a}nchez}, {Stanishev},
  {Miralles-Caballero}, {D{\'\i}az}, {S{\'a}nchez-Blazquez}, {Moll{\'a}},
  {Gallazzi}, {Papaderos}, {Gomes}, {Gruel}, {P{\'e}rez}, {Ruiz-Lara},
  {Florido}, {de Lorenzo-C{\'a}ceres}, {Mendez-Abreu}, {Kehrig}, {Roth},
  {Ziegler}, {Alves}, {Wisotzki}, {Kupko}, {Quirrenbach}, {Bomans}, \& {CALIFA
  Collaboration}}]{Sanchez2013}
{S{\'a}nchez}, S.~F., {Rosales-Ortega}, F.~F., {Jungwiert}, B., {et~al.} 2013,
  \aap, 554, A58, \dodoi{10.1051/0004-6361/201220669}

\bibitem[{{S{\'a}nchez} {et~al.}(2022){S{\'a}nchez}, {Barrera-Ballesteros},
  {Lacerda}, {Mej{\'\i}a-Narvaez}, {Camps-Fari{\~n}a}, {Bruzual},
  {Espinosa-Ponce}, {Rodr{\'\i}guez-Puebla}, {Calette}, {Ibarra-Medel},
  {Avila-Reese}, {Hernandez-Toledo}, {Bershady}, {Cano-Diaz}, \&
  {Munguia-Cordova}}]{Sanchez2022}
{S{\'a}nchez}, S.~F., {Barrera-Ballesteros}, J.~K., {Lacerda}, E., {et~al.}
  2022, \apjs, 262, 36, \dodoi{10.3847/1538-4365/ac7b8f}

\bibitem[{{S{\'a}nchez Almeida} {et~al.}(2014){S{\'a}nchez Almeida},
  {Elmegreen}, {Mu{\~n}oz-Tu{\~n}{\'o}n}, \& {Elmegreen}}]{SanchezAlmeida2014}
{S{\'a}nchez Almeida}, J., {Elmegreen}, B.~G., {Mu{\~n}oz-Tu{\~n}{\'o}n}, C.,
  \& {Elmegreen}, D.~M. 2014, \aapr, 22, 71, \dodoi{10.1007/s00159-014-0071-1}

\bibitem[{{S{\'a}nchez-Menguiano} {et~al.}(2016){S{\'a}nchez-Menguiano},
  {S{\'a}nchez}, {P{\'e}rez}, {Garc{\'\i}a-Benito}, {Husemann}, {Mast},
  {Mendoza}, {Ruiz-Lara}, {Ascasibar}, {Bland-Hawthorn}, {Cavichia},
  {D{\'\i}az}, {Florido}, {Galbany}, {G{\'o}nzalez Delgado}, {Kehrig},
  {Marino}, {M{\'a}rquez}, {Masegosa}, {M{\'e}ndez-Abreu}, {Moll{\'a}}, {Del
  Olmo}, {P{\'e}rez}, {S{\'a}nchez-Bl{\'a}zquez}, {Stanishev}, {Walcher},
  {L{\'o}pez-S{\'a}nchez}, \& {CALIFA Collaboration}}]{SanchezMenguiano2016}
{S{\'a}nchez-Menguiano}, L., {S{\'a}nchez}, S.~F., {P{\'e}rez}, I., {et~al.}
  2016, \aap, 587, A70, \dodoi{10.1051/0004-6361/201527450}

\bibitem[{{Schoenmakers} {et~al.}(1997){Schoenmakers}, {Franx}, \& {de
  Zeeuw}}]{Schoenmakers1997}
{Schoenmakers}, R.~H.~M., {Franx}, M., \& {de Zeeuw}, P.~T. 1997, \mnras, 292,
  349, \dodoi{10.1093/mnras/292.2.349}

\bibitem[{{Scudder} {et~al.}(2012){Scudder}, {Ellison}, {Torrey}, {Patton}, \&
  {Mendel}}]{Scudder2012}
{Scudder}, J.~M., {Ellison}, S.~L., {Torrey}, P., {Patton}, D.~R., \& {Mendel},
  J.~T. 2012, \mnras, 426, 549, \dodoi{10.1111/j.1365-2966.2012.21749.x}

\bibitem[{{Shapiro} {et~al.}(2008){Shapiro}, {Genzel}, {F{\"o}rster Schreiber},
  {Tacconi}, {Bouch{\'e}}, {Cresci}, {Davies}, {Eisenhauer}, {Johansson},
  {Krajnovi{\'c}}, {Lutz}, {Naab}, {Arimoto}, {Arribas}, {Cimatti}, {Colina},
  {Daddi}, {Daigle}, {Erb}, {Hernandez}, {Kong}, {Mignoli}, {Onodera},
  {Renzini}, {Shapley}, \& {Steidel}}]{Shapiro2008}
{Shapiro}, K.~L., {Genzel}, R., {F{\"o}rster Schreiber}, N.~M., {et~al.} 2008,
  \apj, 682, 231, \dodoi{10.1086/587133}

\bibitem[{{Simard} {et~al.}(2011){Simard}, {Mendel}, {Patton}, {Ellison}, \&
  {McConnachie}}]{Simard2011}
{Simard}, L., {Mendel}, J.~T., {Patton}, D.~R., {Ellison}, S.~L., \&
  {McConnachie}, A.~W. 2011, \apjs, 196, 11, \dodoi{10.1088/0067-0049/196/1/11}

\bibitem[{{Smee} {et~al.}(2013){Smee}, {Gunn}, {Uomoto}, {Roe}, {Schlegel},
  {Rockosi}, {Carr}, {Leger}, {Dawson}, {Olmstead}, {Brinkmann}, {Owen},
  {Barkhouser}, {Honscheid}, {Harding}, {Long}, {Lupton}, {Loomis}, {Anderson},
  {Annis}, {Bernardi}, {Bhardwaj}, {Bizyaev}, {Bolton}, {Brewington}, {Briggs},
  {Burles}, {Burns}, {Castander}, {Connolly}, {Davenport}, {Ebelke}, {Epps},
  {Feldman}, {Friedman}, {Frieman}, {Heckman}, {Hull}, {Knapp}, {Lawrence},
  {Loveday}, {Mannery}, {Malanushenko}, {Malanushenko}, {Merrelli}, {Muna},
  {Newman}, {Nichol}, {Oravetz}, {Pan}, {Pope}, {Ricketts}, {Shelden},
  {Sandford}, {Siegmund}, {Simmons}, {Smith}, {Snedden}, {Schneider},
  {SubbaRao}, {Tremonti}, {Waddell}, \& {York}}]{Smee2013}
{Smee}, S.~A., {Gunn}, J.~E., {Uomoto}, A., {et~al.} 2013, \aj, 146, 32,
  \dodoi{10.1088/0004-6256/146/2/32}

\bibitem[{{Sofue} \& {Rubin}(2001)}]{Sofue2001}
{Sofue}, Y., \& {Rubin}, V. 2001, \araa, 39, 137,
  \dodoi{10.1146/annurev.astro.39.1.137}

\bibitem[{{Somerville} \& {Dav{\'e}}(2015)}]{Somerville2015}
{Somerville}, R.~S., \& {Dav{\'e}}, R. 2015, \araa, 53, 51,
  \dodoi{10.1146/annurev-astro-082812-140951}

\bibitem[{{Spekkens} \& {Sellwood}(2007)}]{Spekkens2007}
{Spekkens}, K., \& {Sellwood}, J.~A. 2007, \apj, 664, 204,
  \dodoi{10.1086/518471}

\bibitem[{{Stark} {et~al.}(2021){Stark}, {Masters}, {Avila-Reese}, {Riffel},
  {Riffel}, {Boardman}, {Zheng}, {Weijmans}, {Dillon}, {Fielder}, {Finnegan},
  {Fofie}, {Goddy}, {Harrington}, {Pace}, {Rujopakarn}, {Samanso}, {Shamsi},
  {Sharma}, {Warrick}, {Witherspoon}, \& {Wolthuis}}]{Stark2021}
{Stark}, D.~V., {Masters}, K.~L., {Avila-Reese}, V., {et~al.} 2021, \mnras,
  503, 1345, \dodoi{10.1093/mnras/stab566}

\bibitem[{{Stewart} {et~al.}(2011){Stewart}, {Kaufmann}, {Bullock}, {Barton},
  {Maller}, {Diemand}, \& {Wadsley}}]{Stewart2011}
{Stewart}, K.~R., {Kaufmann}, T., {Bullock}, J.~S., {et~al.} 2011, \apj, 738,
  39, \dodoi{10.1088/0004-637X/738/1/39}

\bibitem[{{Tacchella} {et~al.}(2016){Tacchella}, {Dekel}, {Carollo},
  {Ceverino}, {DeGraf}, {Lapiner}, {Mandelker}, \& {Primack
  Joel}}]{Tacchella2016}
{Tacchella}, S., {Dekel}, A., {Carollo}, C.~M., {et~al.} 2016, \mnras, 457,
  2790, \dodoi{10.1093/mnras/stw131}

\bibitem[{{Torrey} {et~al.}(2018){Torrey}, {Vogelsberger}, {Hernquist},
  {McKinnon}, {Marinacci}, {Simcoe}, {Springel}, {Pillepich}, {Naiman},
  {Pakmor}, {Weinberger}, {Nelson}, \& {Genel}}]{Torrey2018}
{Torrey}, P., {Vogelsberger}, M., {Hernquist}, L., {et~al.} 2018, \mnras, 477,
  L16, \dodoi{10.1093/mnrasl/sly031}

\bibitem[{{Tremonti} {et~al.}(2004){Tremonti}, {Heckman}, {Kauffmann},
  {Brinchmann}, {Charlot}, {White}, {Seibert}, {Peng}, {Schlegel}, {Uomoto},
  {Fukugita}, \& {Brinkmann}}]{Tremonti2004}
{Tremonti}, C.~A., {Heckman}, T.~M., {Kauffmann}, G., {et~al.} 2004, \apj, 613,
  898, \dodoi{10.1086/423264}

\bibitem[{{Tumlinson} {et~al.}(2017){Tumlinson}, {Peeples}, \&
  {Werk}}]{Tumlinson2017}
{Tumlinson}, J., {Peeples}, M.~S., \& {Werk}, J.~K. 2017, \araa, 55, 389,
  \dodoi{10.1146/annurev-astro-091916-055240}

\bibitem[{{van der Kruit} \& {Allen}(1978)}]{vanderKruit1978}
{van der Kruit}, P.~C., \& {Allen}, R.~J. 1978, \araa, 16, 103,
  \dodoi{10.1146/annurev.aa.16.090178.000535}

\bibitem[{{Veilleux} {et~al.}(2005){Veilleux}, {Cecil}, \&
  {Bland-Hawthorn}}]{Veilleux2005}
{Veilleux}, S., {Cecil}, G., \& {Bland-Hawthorn}, J. 2005, \araa, 43, 769,
  \dodoi{10.1146/annurev.astro.43.072103.150610}

\bibitem[{{Venturi} {et~al.}(2018){Venturi}, {Nardini}, {Marconi}, {Carniani},
  {Mingozzi}, {Cresci}, {Mannucci}, {Risaliti}, {Maiolino}, {Balmaverde},
  {Bongiorno}, {Brusa}, {Capetti}, {Cicone}, {Ciroi}, {Feruglio}, {Fiore},
  {Gallazzi}, {La Franca}, {Mainieri}, {Matsuoka}, {Nagao}, {Perna},
  {Piconcelli}, {Sani}, {Tozzi}, \& {Zibetti}}]{Venturi2018}
{Venturi}, G., {Nardini}, E., {Marconi}, A., {et~al.} 2018, \aap, 619, A74,
  \dodoi{10.1051/0004-6361/201833668}

\bibitem[{{Wake} {et~al.}(2017){Wake}, {Bundy}, {Diamond-Stanic}, {Yan},
  {Blanton}, {Bershady}, {S{\'a}nchez-Gallego}, {Drory}, {Jones}, {Kauffmann},
  {Law}, {Li}, {MacDonald}, {Masters}, {Thomas}, {Tinker}, {Weijmans}, \&
  {Brownstein}}]{Wake2017}
{Wake}, D.~A., {Bundy}, K., {Diamond-Stanic}, A.~M., {et~al.} 2017, \aj, 154,
  86, \dodoi{10.3847/1538-3881/aa7ecc}

\bibitem[{{Walmsley} {et~al.}(2023){Walmsley}, {G{\'e}ron}, {Kruk}, {Scaife},
  {Lintott}, {Masters}, {Dawson}, {Dickinson}, {Fortson}, {Garland}, {Mantha},
  {O'Ryan}, {Popp}, {Simmons}, {Baeten}, \& {Macmillan}}]{Walmsley2023}
{Walmsley}, M., {G{\'e}ron}, T., {Kruk}, S., {et~al.} 2023, \mnras, 526, 4768,
  \dodoi{10.1093/mnras/stad2919}

\bibitem[{{Wang} \& {Lilly}(2023)}]{Wang2023}
{Wang}, E., \& {Lilly}, S.~J. 2023, \apj, 944, 143,
  \dodoi{10.3847/1538-4357/acaf31}

\bibitem[{{Westfall} {et~al.}(2019){Westfall}, {Cappellari}, {Bershady},
  {Bundy}, {Belfiore}, {Ji}, {Law}, {Schaefer}, {Shetty}, {Tremonti}, {Yan},
  {Andrews}, {Brownstein}, {Cherinka}, {Coccato}, {Drory}, {Maraston},
  {Parikh}, {S{\'a}nchez-Gallego}, {Thomas}, {Weijmans}, {Barrera-Ballesteros},
  {Du}, {Goddard}, {Li}, {Masters}, {Ibarra Medel}, {S{\'a}nchez}, {Yang},
  {Zheng}, \& {Zhou}}]{Westfall2019}
{Westfall}, K.~B., {Cappellari}, M., {Bershady}, M.~A., {et~al.} 2019, \aj,
  158, 231, \dodoi{10.3847/1538-3881/ab44a2}

\bibitem[{{Wong} {et~al.}(2004){Wong}, {Blitz}, \& {Bosma}}]{Wong2004}
{Wong}, T., {Blitz}, L., \& {Bosma}, A. 2004, \apj, 605, 183,
  \dodoi{10.1086/382215}

\bibitem[{{Yan} {et~al.}(2016{\natexlab{a}}){Yan}, {Bundy}, {Law}, {Bershady},
  {Andrews}, {Cherinka}, {Diamond-Stanic}, {Drory}, {MacDonald},
  {S{\'a}nchez-Gallego}, {Thomas}, {Wake}, {Weijmans}, {Westfall}, {Zhang},
  {Arag{\'o}n-Salamanca}, {Belfiore}, {Bizyaev}, {Blanc}, {Blanton},
  {Brownstein}, {Cappellari}, {D'Souza}, {Emsellem}, {Fu}, {Gaulme}, {Graham},
  {Goddard}, {Gunn}, {Harding}, {Jones}, {Kinemuchi}, {Li}, {Li}, {Maiolino},
  {Mao}, {Maraston}, {Masters}, {Merrifield}, {Oravetz}, {Pan}, {Parejko},
  {Sanchez}, {Schlegel}, {Simmons}, {Thanjavur}, {Tinker}, {Tremonti}, {van den
  Bosch}, \& {Zheng}}]{Yan2016b}
{Yan}, R., {Bundy}, K., {Law}, D.~R., {et~al.} 2016{\natexlab{a}}, \aj, 152,
  197, \dodoi{10.3847/0004-6256/152/6/197}

\bibitem[{{Yan} {et~al.}(2016{\natexlab{b}}){Yan}, {Tremonti}, {Bershady},
  {Law}, {Schlegel}, {Bundy}, {Drory}, {MacDonald}, {Bizyaev}, {Blanc},
  {Blanton}, {Cherinka}, {Eigenbrot}, {Gunn}, {Harding}, {Hogg},
  {S{\'a}nchez-Gallego}, {S{\'a}nchez}, {Wake}, {Weijmans}, {Xiao}, \&
  {Zhang}}]{Yan2016a}
{Yan}, R., {Tremonti}, C., {Bershady}, M.~A., {et~al.} 2016{\natexlab{b}}, \aj,
  151, 8, \dodoi{10.3847/0004-6256/151/1/8}

\bibitem[{{Yu} {et~al.}(2021){Yu}, {Ho}, \& {Wang}}]{Yu2021}
{Yu}, S.-Y., {Ho}, L.~C., \& {Wang}, J. 2021, \apj, 917, 88,
  \dodoi{10.3847/1538-4357/ac0c77}

\end{thebibliography}
\bibliographystyle{aasjournal}

\end{CJK*}
\end{document}